\documentclass[10pt,journal,cspaper,compsoc]{IEEEtran} % TDSC
\usepackage{url,graphicx,amssymb,amsfonts}
\usepackage{geometry}
\usepackage[flushleft]{threeparttable}
\usepackage[pdfstartview=FitH,colorlinks, citecolor=blue]{hyperref}
\usepackage{amsmath}
\usepackage{tikz}
\usepackage{algorithm}
\usepackage{algpseudocode}
\usepackage[english]{babel}
\usepackage{blindtext}
\usepackage{multicol}
\usepackage[font=small,labelfont=bf]{caption}

\usetikzlibrary{automata, positioning, arrows}
\tikzset{->, initial text=$ $}

\makeatletter
\def\BState{\State\hskip-\ALG@thistlm}
\makeatother
\usepackage{amsthm} % define theorem, lemma, proof
\newtheorem{theorem}{Theorem}[section]

\newtheorem{claim}[theorem]{Claim}

\newtheorem{definition}[theorem]{Definition}

\newcommand{\Fvote}{\mathcal{F}_{\textsc{Vote}}^{n,k,t,\delta}}

\newcommand{\Fbb}{\mathcal{G}_{\textsc{BB}}}

\newcommand{\Fkeygen}{\mathcal{F}_{\textsc{KeyGen}}}

\newcommand{\Fdec}{\mathcal{F}_{\textsc{Dec}}}

\newcommand{\mFcert}{\hat{\mathcal{G}}_{\textsc{Cert}}}

\newcommand{\Fvsd}{\mathcal{F}_{\textsc{VSD}}}

\newcommand{\Fasd}{\mathcal{F}_{\textsc{ASD}}}

\newcommand{\Fve}{\mathcal{F}_{\textsc{V}.\textsc{Emu}}}
\newcommand{\Voter}{\mathsf{V}}

\newcommand{\AU}{\mathsf{AU}}
\newcommand{\Trustee}{\mathsf{T}}

\newcommand{\Ext}{\mathsf{Ext}}

\newcommand{\Complain}{\mathsf{Complain}}

\newcommand{\Pivote}{\Pi_{\mathsf{vote}}}

\newcommand{\Enc}{\mathsf{Enc}}

\newcommand{\Pub}{\mathsf{Pub}}

\newcommand{\Priv}{\mathsf{Priv}}

\newcommand{\Audited}{\textsc{Audited}}

\newcommand{\End}{\textsc{End}}

\newcommand{\Begin}{\textsc{Begin}}

\newcommand{\Ballots}{\mathsf{Ballots}}

\newcommand{\Alert}{\mathsf{Alert}}
\newcommand{\Tamper}{\textsc{Tamper}}

\newcommand{\cor}{\mathsf{cor}}

\newcommand{\HW}{\mathsf{HW}}

\newcommand{\Read}{\textsc{Read}}

\newcommand{\Corrupt}{\textsc{Corrupt}}

\newcommand{\Start}{\textsc{Start}}
\newcommand{\Vote}{\textsc{Vote}}
\newcommand{\Tally}{\textsc{Tally}}
\newcommand{\Audit}{\textsc{Audit}}
\newcommand{\Proceed}{\textsc{Proceed}}

\newcommand{\PubPost}{\textsc{PubPost}}

\newcommand{\PrivPost}{\textsc{PrivPost}}

\newcommand{\TallyAlg}{\mathsf{TallyAlg}}

\newcommand{\Leak}{\textsc{Leak}}

\newcommand{\PrivKey}{\textsc{PrivKey}}
\newcommand{\PubKey}{\textsc{PubKey}}
\newcommand{\Key}{\textsc{Key}}

\newcommand{\Result}{\textsc{Result}}

\newcommand{\Emulate}{\textsc{Emulate}}

\newcommand{\Ready}{\textsc{Ready}}

\newcommand{\TM}{\mathsf{M}}

\newcommand{\TMH}{\mathsf{M}_{\textsc{Honest}}}

\newcommand{\status}{\mathsf{status}}

\newcommand{\Sign}{\textsc{Sign}}

\newcommand{\Signature}{\textsc{Signature}}

\newcommand{\SignEnc}{\textsc{SignEnc}}

\newcommand{\Ballot}{\textsc{Ballot}}

\newcommand{\Receipt}{\textsc{Receipt}}

\newcommand{\Act}{\mathsf{script}}

\newcommand{\Record}{\mathsf{Record}}

\newcommand{\exec}{\mathsf{Exec}}

\newcommand{\PK}{\mathsf{PK}}
\newcommand{\SK}{\mathsf{SK}}

\newcommand{\KeyGen}{\mathsf{KeyGen}}

\newcommand{\NIZK}{\mathsf{NIZK}}

\newcommand{\Dec}{\mathsf{Dec}}

\newcommand{\Rand}{\mathsf{Rand}}

\newcommand{\TDec}{\mathsf{TDec}}

\newcommand{\Reconstruct}{\mathsf{Reconstruct}}

\newcommand{\Deal}{\mathsf{Deal}}

\newcommand{\invalid}{\mathsf{invalid}}
\newcommand{\valid}{\mathsf{valid}}

\newcommand{\Ver}{\mathsf{Ver}}
\newcommand{\Sim}{\mathsf{Sim}}

\newcommand{\Prov}{\mathsf{Prov}}

\newcommand{\poly}{\mathsf{poly}}
\newcommand{\negl}{\mathsf{negl}}

\newcommand{\NN}{\mathbb{N}}
\newcommand{\ZZ}{\mathbb{Z}}

\newcommand{\AAA}{\mathcal{A}}
\newcommand{\BBB}{\mathcal{B}}

\newcommand{\DDD}{\mathcal{D}}

\newcommand{\LLL}{\mathcal{L}}
\newcommand{\FFF}{\mathcal{F}}
\newcommand{\HHH}{\mathcal{H}}

\newcommand{\SSS}{\mathcal{S}}
\newcommand{\KKK}{\mathcal{K}}
\newcommand{\TTT}{\mathcal{T}}

\newcommand{\ZZZ}{\mathcal{Z}}

\newcommand{\RRR}{\mathcal{R}}

\newcommand{\VVV}{\mathcal{V}}

\newcommand{\JJJ}{\mathcal{J}}

\usepackage{longtable,booktabs}
\usepackage{adjustbox} % \linewidth or \columnwidth
\usepackage{mathtools}

\definecolor{lgray}{gray}{0.5}

\newcommand{\secp}{\ensuremath{\lambda}}

\newcommand{\ct}{\ensuremath{ct}}

\newcommand{\PPP}{\ensuremath{\mathcal{P}}}

\newcommand{\Setup}{\ensuremath{\mathsf{Setup}}}

\renewcommand{\Tally}{\ensuremath{\mathrm{Tally}}}
\renewcommand{\Result}{\ensuremath{\mathrm{Result}}}
\newcommand{\Verify}{\ensuremath{\mathrm{Verify}}}
\newcommand{\EA}{\ensuremath{\textsf{EA}}}
\newcommand{\BB}{\ensuremath{\textsf{BB}}\xspace}
\newcommand{\ASD}{\ensuremath{\textsf{ASD}}}

\newcommand{\VSD}{\ensuremath{\textsf{VSD}}}

\newcommand{\VOTE}{\ensuremath{\mathsf{VOTE}}}

\newcommand{\params}{\ensuremath{\mathsf{params}}}
\newcommand{\st}{\ensuremath{\mathsf{st}}}

\newcommand{\sid}{\ensuremath{sid}}
\newcommand{\ssid}{\ensuremath{ssid}}

\newcommand{\Z}{\ensuremath{\mathbb{Z}}}

\newcommand{\Zq}{\ensuremath{\Z_q}}

\newcommand{\msg}{\ensuremath{m}}
\newcommand{\mybox}[5]{
    \begin{figure}[tp!]%[htb!]%[tp]
        \centering
    \begin{tikzpicture}
        \node[anchor=text,text width=\columnwidth-1cm, draw, rounded corners, line width=1pt, fill=#3, inner sep=2mm] (big) {\\#4};
        \node[draw, rounded corners, line width=.5pt, fill=#2, anchor=west, xshift=5mm] (small) at (big.north west) {#1};
    \end{tikzpicture}
    \caption{#5}
    \end{figure}
}

\newcommand{\myfullbox}[5]{
    \begin{figure*}[tp!]%[htb!]%[tp]
        \centering
    \begin{tikzpicture}
        \node[anchor=text,text width=2*\columnwidth-1cm, draw, rounded corners, line width=1pt, fill=#3, inner sep=2mm] (big) {\\#4};
        \node[draw, rounded corners, line width=.5pt, fill=#2, anchor=west, xshift=5mm] (small) at (big.north west) {#1};
    \end{tikzpicture}
    \caption{#5}
    \end{figure*}
    \vspace{-2mm}
}

\usepackage{setspace}
\usepackage{enumitem}
\usepackage{cancel}
\usepackage[normalem]{ulem}
\newcommand{\delete}[1]{{\color{lgray}{\sout{#1}}}\xspace}

\begin{document}
%


\title{UC Modelling and Security Analysis of the Estonian IVXV Internet Voting System}
%%\erhao{\textsf{PPTauth}
%To Guess or Not to Guess: Two-Factor
%%  User Corruption   and Server Compromise


\author{Bingsheng Zhang, Zengpeng Li, \and Jan Willemson
\IEEEcompsocitemizethanks{
\IEEEcompsocthanksitem B. Zhang is with the School of Cyber Science and Technology, Zhejiang University,  Hangzhou, P.R. China .(e-mail: bingsheng@zju.edu.cn).
\IEEEcompsocthanksitem Z. Li is the corresponding author and he is with the School of Cyber Science and Technology, Shandong University, Qingdao, P.R. China. (Email: zengpeng@email.sdu.edu.cn).
\IEEEcompsocthanksitem J. Willemson is with the Cybernetica, Ulikooli, Tartu, Estonia.
 (E-mail: janwil@cyber.ee.)
}}
%\maketitle % TIFS position
%\maketitle
%--TDSC-start---
\IEEEcompsoctitleabstractindextext{
%! TEX encoding =UTF-8 Unicode
%! TEX root = mainArxiv.tex

%\vspace{-15pt}
\begin{abstract}
Estonian Internet voting has been used in national-wide elections since 2005. However, the system was initially designed in a \emph{heuristic manner}, with very few proven security guarantees. The Estonian Internet voting system has constantly been evolving throughout the years, with the latest version (code-named IVXV) implemented in 2018. Nevertheless, to date, no formal security analysis of the system has been given.
In this work, for the first time, we provide a rigorous security modeling for the Estonian IVXV system as a \emph{ceremony}, attempting to capture the effect of actual human behavior on election verifiability in the universal composability (UC) framework.
Based on the voter behavior statistics collected from three actual election events in Estonia, we show that IVXV achieves end-to-end verifiability in practice despite the fact that only $4\%$ (on average) of the Estonian voters audit their ballots.

\end{abstract}
\begin{IEEEkeywords}
 E-voting Ceremony; Universal Composability; End-to-End Verifiability; Estonian IVXV Internet Voting
\end{IEEEkeywords}

%
%
%Estonian Internet voting  is the world's first remote voting system that has been used in national-wide elections. To date, Estonia remains the only country allowing access to Internet voting to all of its citizens. However, the system was initially designed in an engineering manner, and it has several security vulnerabilities. Throughout the years, the Estonian Internet voting system is constantly evolving, and the latest version is code-named IVXV. Nevertheless, there is no formal systematic security analysis of the IVXV system.
%
%In this work, we aim to put a foundation on abstracting, modelling and analysing the security of IVXV. A verifiable e-voting scheme usually requires the voters, beyond the ballot-casting procedure, to carry out additional auditing steps for individual verifiability. Voter's defective execution of such auditing steps may affect the soundness of the election result. We collected Estonian real voter behaviour statistics for the recent 5-year legally binding national-wide elections, and study IVXV as an e-voting ceremony, where the voters' behaviour are taken into account in the security analysis.
%In particular, for the first time, we provide a rigorous security modelling for an e-voting ceremony, attempting to capture human actions in the  well-known UC framework. We show that IVXV achieves end-to-end verifiability in practice even less than 4\% of the voters audit their ballots.

}
\maketitle % TDSC position
%%--TDSC-end---
%%--TIFS-start---
%\input{abstract}
%\textbf{\emph{keywords--} E-voting Ceremony; Universal Composability; End-to-End Verifiability; Estonian IVXV Internet Voting.}
%%--TIFS-end---
% paper start
\setlength{\emergencystretch}{3pt}
\section{Introduction}
%\vspace{-5pt}
%! TEX encoding =UTF-8 Unicode
%!TEX root = mainTDSC20.tex
In recent years, the advancement of remote electronic voting is gradually moving from an academic, mental exercise into practice.
The first Internet voting events with legally binding results were conducted already in 2000 in Arizona~\cite{NSFInternetVoting} and the University of Osnabr\"{u}ck, Germany~\cite{LindenauMSc}. Only five years later, in 2005, the option of vote casting over the Internet was provided in the country-wide elections in Estonia for the first time in the world~\cite{MaMa06}. By 2014, the share of Internet votes in Estonia reached over 30\%~\cite{DBLP:conf/voteid/VinkelK16}.
There have been various attempts to set up remote Internet voting in other countries, such as Switzerland~\cite{DBLP:conf/ev/Driza-MaurerSTW12} and Norway~\cite{DBLP:conf/ev/StenerudB12}, but to date, Estonia remains the only country allowing access to Internet voting to all of its citizens.

While developing its Internet voting solution, Estonia has taken an engineering rather than theoretically well-founded approach. The {initial protocol} has been kept relatively simple, mimicking the conventional double-envelope postal voting and making use of the strong eID frameworks (ID-card and Mobile-ID) provided to the citizens. After its launch, the Estonian Internet voting system has been evolving towards better security guarantees.
For instance, in 2011, a student presented a proof-of-concept vote manipulation malware making use of the fact that at that time, there was no vote verification mechanism implemented in the Estonian i-vote system~\cite{DBLP:conf/voteid/HeibergLW11}. By the next election event in 2013, the option of verifying the integrity of the votes stored on the server using a personal mobile device was introduced~\cite{HW11}.
In contrast, the server-side of the Estonian Internet voting system was only protected using physical and organizational measures, not verifiable by the general public in an irrefutable manner. Several problems occurring as a result of such an architecture were pointed out by Springall \emph{et al.} in 2014~\cite{springall2014security}. To address these problems, the central system architecture was entirely redesigned by the 2017 parliamentary elections, introducing a mix-net procedure for vote privacy, provable decryption of votes, and independent vote commitments to ensure digital ballot box integrity~\cite{heiberg2016improving}. The resulting system is code-named IVXV~\cite{IVXV}.

However, to date, there is no formal systematic security analysis of the IVXV system. In this work, we aim to put a foundation for abstract, model, and analyze the security of the IVXV system. In particular, we are interested in \emph{end-to-end (E2E) verifiability}, which has been widely accepted as a fundamental requirement for e-voting adoption.
In an end-to-end verifiable e-voting system, the voter can obtain a receipt after ballot casting, allowing him/her to verify that his/her vote was (i) cast as intended, (ii) recorded as cast, and (iii) tallied as recorded. Moreover, any external third-party auditor should also be able to verify that the election procedure was executed properly.
Unfortunately, merely showing that a system is E2E verifiable does not imply the integrity of the election result in the real world. This is because the E2E verifiability usually relies on human participant behavior in a highly non-trivial manner. The ability of human voters to compromise overall security due to their negligence is well known in e-voting system design  (cf.~\cite{KSW05}). For instance, in terms of individual verifiability, to achieve the ``cast-as-intended'' property, the voters are typically required/advised to perform specific auditing.
This means that the voters, beyond the ballot-casting procedure, are
supposed to carry out additional steps that many may find to be counterintuitive,
see e.g., \cite{Volkamer13} for more %\change{more} \remove{a}
discussion of this issue. This potentially leads to the defective execution
of the appropriate steps that are to be carried out for verifiability to be supported; therefore, the verifiability of the election may collapse.
In fact, according to the Estonian Internet voting statistics collected from the recent 5-year legally binding national-wide elections, only approximately  $4\%$ of voters perform the recommended ballot auditing process~\cite{DBLP:conf/voteid/HeibergPW15}.

Is the Estonian IVXV system E2E verifiable w.r.t. the Estonian voter profile in practice?  To answer this question, we study IVXV security as an \emph{e-voting ceremony}. The notion, \emph{ceremony}, was introduced by Ellison \cite{DBLP:journals/iacr/Ellison07} to expand a security protocol with out-of-band channels, and the human users are considered as separate nodes of the system that should be taken into account when performing the security analysis.
Later, the `conditioned-safe ceremony' notion that encompasses forcing functions, defense-in-depth, and human tendencies was introduced in~\cite{ceremony09}.  In 2015, Johansen and J{\o}sang~\cite{ceremony15} proposed a formal probabilistic model for verifying a security ceremony. In their work, the human agent interaction with the user interface is modeled as a non-deterministic process. In 2017, Kiayias \emph{et al.}~\cite{DBLP:conf/pkc/Kiayias0Z17} analyzed the security of Helios in terms of an e-voting ceremony using property-based definitions. In this work, we, for the first time, model an e-voting ceremony in the
\emph{universal composability} (UC) framework~\cite{FOCS:Canetti01,TCC:CDPW07}.

%To the best of our knowledge, this is the first attempt also to capture human actions in the universal composability framework.
\vspace{3pt}
\noindent\textbf{Our contributions.}
%\subsection{Our contributions}
%
(i) For the first time, we provide a rigorous security modeling
for an e-voting ceremony, attempting to capture human actions in the  well-known UC (generalized UC, a.k.a. gUC)  framework~\cite{FOCS:Canetti01,TCC:CDPW07}.
To model verifiability, we introduce a conceptual entry, auditor $\AU$, which cannot be corrupted.\footnote{In the real world, there are in general several auditors who study the evidence presented to them independently. It is assumed that all of them should come to a coordinated decision concerning the correctness and consistency of this evidence. However, this process remains outside of the cryptographic realm, and hence we model $\AU$ as one entity here.} The adversary is allowed to tamper with $O(\log \secp)$ ballots without raising an alert, i.e., the difference between the announced tally and the true tally is logarithmically bounded w.r.t. the security parameter $\lambda$.\footnote{Since the number of voters $n=\poly(\lambda)$, such a difference can also be viewed as $O(\log n)$. This is used to model a small tally deviation that is not sufficient to swing the election results in practice. For an election with several billion voters, the deviation could remain as low as hundreds.}
  When $\omega(\secp)$ ballots are tampered with, the $\AU$ will return $\invalid$ and void the entire election. In the spirit of e-voting ceremony, we distinguish the (human) voters from their \emph{voter supporting devices} ($\VSD$s), such as PC, smartphone, tablet, etc. This separation enables more refined security analysis, e.g., covering the case when a voter is honest but his/her $\VSD$ is compromised. The $\VSD$ of voter $\Voter_i$ is modeled as an ideal functionality $\Fvsd^{\Voter_i}$. $\Fvsd^{\Voter_i}$ is parameterized with a PPT Turing machine $\TMH$, which is used to model the software that is running on the $\VSD$. The functionality of $\TMH$ is encrypting and signing the voter's ballots. The adversary is able to compromise the $\Fvsd^{\Voter_i}$ by sending an arbitrary
$\TM$ to it, and $\Fvsd^{\Voter_i}$ will use the modified $\TM$ for all its operations. Note that if the $\Fvsd^{\Voter_i}$ is compromised, the choice of voter $\Voter_i$ is leaked to the adversary.
 In addition, \emph{auditing supporting devices} ($\ASD$s) are introduced to allow the voters to have a trustworthy device to perform certain ballot auditing operations.
 Although each voter may have its own $\ASD$, the corresponding operation is universal. Hence we use a single functionality $\Fasd$ to model all the $\ASD$s in our model. %$\Fasd$ is always trusted in our model. %nevertheless, the Estonian IVXV system remain verifiable even any constant fraction of the $\ASD$ are compromised.
 %(See Sec.~\ref{sec:securitymodel} for details.)

  We assume that the voters as human beings are not capable of performing complicated cryptographic operations or generating high entropy randomness. (In fact, in this work, the voters are modeled as finite state machines, but they are only required to perform string forwarding and matching tasks.) As each voter may have a different behavioral pattern in practice, we introduce a special voter emulator functionality $\Fve^\DDD$ that is parameterized by a voter profile distribution $\DDD$ and sample a voter behavior pattern according to $\DDD$ as a script. The voter then runs the script during the election.

\vspace{3pt}
(ii) For the first time, we formally abstract the latest Estonian IVXV system as described in \cite{IVXV,IVXVgit}
in terms of a UC protocol. More specifically, we introduce several ideal functionaries, i.e., $\Fbb, \mFcert,\{\Fvsd^{\Voter_i}\}_{i\in[n]}, \Fasd, \Fkeygen, \Fdec$ and $\Fve^\DDD$, to abstract the implementation details away from the main protocol description. We then discuss how each ideal functionality is instantiated in practice and list the implied assumptions in Sec.~\ref{sec:implementation}. For instance, the multi-session certificate functionality $\mFcert$ is introduced to abstract the Estonian PKI infrastructure. It provides a direct binding between a signature and the identity of the corresponding signer. Note that the bulletin board (BB) functionality $\Fbb$ is slightly different from the conventional BB functionality required by an end-to-end verifiable voting system. Normally, a BB is publicly accessible in the sense that anyone can read the messages posted on the BB. The $\Fbb$ used in the Estonian IVXV system consists of public BB and private BB, where the public BB is the same as conventional BB, but the private BB can only be accessed by the auditor(s) and $\ASD$s. This setup can enable verifiability while still achieving a certain level of coercion resistance by preventing the adversary from seeing certain BB content\footnote{In particular, the sensitive information includes the time-stamps of encrypted and signed votes. The Estonian system provides the voters with the option of re-voting in case they were forced to cast a vote against their true will. However, the efficiency of this measure relies on the coercer not being able to understand which one of the votes submitted by the voter was the latest. If the attacker is sufficiently powerful to obtain the timing information, he can request the voter to reveal her verification token and complete the coercive attack (as described e.g., in the OSCE/ODIHR report~\cite{OSCE2019}).  
%However, this attack relies on a rather strong assumption of the auditor(s) being malicious. In practice, this threat is mitigated using organizational measures, including background checks of the potential auditors.
}.

\vspace{3pt}
(iii) We build a real-world Estonian voter behavior profile based on the actual statistics of the voting and verification patterns observed during three recent election events in the past five years.
This allows us to analyze the security of the Estonian IVXV system under the proposed UC framework w.r.t. the real Estonian voter behavior distribution. We show that the Estonian IVXV protocol UC-realizes the ideal verifiable voting functionality $\Fvote$, where $n$ is the number of voters, $k$ is the number of trustees, and $t-1$ is the maximum number of corrupted trustees that the protocol can tolerate. We also study the optimal strategy for the adversary to tamper with the votes and estimate the probability of success for this strategy. For example, when the adversary tries to change 100 or 200 random votes, his probability of success is $1.687\%$ or $0.028\%$, respectively.

%\input{related}
%! TEX encoding =UTF-8 Unicode
%! TEX root = mainArxiv.tex

\vspace{3pt}
\noindent\textbf{Roadmap.}
%\subsection{Roadmap}
%The remainder of this paper is organized as follows.
In Sec.~\ref{sec:preliminaries} we formally present notations that will be used throughout the paper. In Sec.~\ref{sec:securitymodel} we describe the ceremony Estonian voting system under the UC framework which contains the ideal functionalities.
In Sec.~\ref{sec:protocol} we describe the ceremony Estonian voting protocol leveraging the proposed ideal functionalities along with security analysis.  Further, in Sec.~\ref{sec:implementation}, we discuss these ideal functionalities are instantiated in the real-world Estonian IVXV system.
Finally, in Sec.~\ref{sec:relatedUC} and Sec.~\ref{sec:conclusion}, we discuss the related works and give a conclusion, respectively. 
%\vspace{-10pt}
\section{Preliminaries}\label{sec:preliminaries}
%! TEX encoding =UTF-8 Unicode
%! TEX root = mainArxiv.tex

%
%%\vspace{-10pt}
%\subsection{Notations}
%%\vspace{3pt}
%%\noindent\textbf{Notations.}
%%Throughout this paper, we use the following notations and terminologies.
%%
%Let $\secp \in\NN$ be the security parameter. Denote the set $\set{a,a+1,\ldots, b}$ by $[a,b]$, let $[b]$ denote $[1,b]$, and let empty set denote $\emptyset$.
%Denote the symmetric group consisting of all the permutations of the set $[N]=\{1,\ldots,N\}$ as $\Sigma_N$. Let $\poly(x)$ and $\negl(x)$ stand for polynomial and negligible quantity in $x$, respectively. We abbreviate \emph{probabilistic polynomial time} as PPT.
%

%
%
%\begin{definition}[Symmetrical Group,\cite{Sgroup07}]
%%Let $S_n$ be a symmetrical group on set $[n]$.
%The sysmetric group $S_n$ is the group of permutations on $n$ objects are labeled $\set{1,2,\cdots,n}$, and elements of $S_n$ are given by bijective functions $\sigma:\set{1,2,\cdots,n}\rightarrow \set{1,2,\cdots,n}$.　The group operation on $S_n$ is composition of functions.
%
%\end{definition}

%\vspace{3pt}
%\noindent\textbf{Threshold secret sharing.}
\subsection{Threshold secret sharing}
The Estonian IVXV voting system adopts Shamir secret sharing~\cite{CACM:Shamir79} for the private key distribution during the election key generation phase. In a $(t,k)$-threshold secret sharing scheme, the secret $x$ is shared by  a dealing algorithm $(s_1,\ldots,s_k)\gets\Deal(x)$. Any group of $t$ or more shares together can open the secret via the reconstruction algorithm $x\gets\Reconstruct(\KKK)$ for $|\KKK|\geq t$ and $\KKK\subseteq \{s_1,\ldots, s_k\}$.

%\vspace{3pt}
%\noindent\textbf{Re-randomizable public key encryption.}
\subsection{Re-randomizable public key encryption}
A public key encryption scheme consists of a tuple of PPT algorithms $\mathsf{PKE}:=(\Setup, \KeyGen, \Enc,\Dec, \Rand)$.

\begin{itemize}
\item $\params\gets \Setup(1^\secp)$: it takes as input the security parameter $\secp$, and outputs a public parameter $\params$.
\item  $(\PK,\SK)\gets\KeyGen(\params)$: it takes as input the public parameter $\params$, and outputs a pair of public key and private key $(\PK,\SK)$.
%If a pair of $(\PK,\SK)$ is generated by $\KeyGen$, we write $(\PK,\SK)\in\RRR_\PK$.

\item $\ct\gets\Enc(\PK,\msg;r)$: it takes as input the public key $\PK$, the message $\msg$, and the randomness $r$, and it outputs the ciphertext $\ct$.

\item $\msg\gets\Dec(\SK, \ct)$: it takes as input the private key $\SK$ and  ciphertext $\ct$, and outputs the message $\msg$.

\item $\ct'\leftarrow\Rand(\PK,\ct ; r)$: it takes as input the public key $\PK$, a ciphertext $\ct$ and the randomness $r$, and outputs a re-randomized ciphertext $\ct'$.

%\item $(\SK_1,\cdots,\SK_n)\gets\Deal(\SK)$: it takes as input the private key and distributes the $\SK$ among participants, it then outputs $(\SK_1,\cdots,\SK_n)$ where each $\SK_i$ allocates to each $P_i$ of participants.
%\item $\SK'\gets\Reconstruct(\KKK)$: it takes as input the set $\KKK$, where $\KKK$ satisfies $\KKK=\{\SK_j:j\in \Gamma\}$ and $\Gamma\subset 2^{\PPP}$, it then outputs $\SK'$.
\end{itemize}

We assume the public key encryption scheme is IND-CPA secure.
In addition, given the randomness used during encryption, we want the ciphertext to be extractable such that there exists an algorithm $\msg\gets\TDec(\PK,\ct, r)$, where $\ct= \Enc(\PK,\msg;r)$.

%\vspace{3pt}
%\noindent\textbf{Non-interactive zero-knowledge proofs.}
\subsection{Non-interactive Zero-Knowledge Proofs}
Let $\RRR$ be an efficiently computable binary relation. For pairs $(x,w) \in \RRR$, we call $x$
the \emph{statement} and $w$ the \emph{witness}. Let $\LLL_\RRR$ be the language
consisting of statements in $\RRR$, i.e. $\LLL_\RRR=\{x|\exists w
\,\, {\rm s.t. } \,\, (x,w)\in \RRR \}$. An non-interactive zero-knowledge (NIZK) proof  consists of PPT algorithms $(\Prov,\Ver,\Sim,\Ext)$, where $\Prov$ is the prover algorithm, $\Ver$ is the verification algorithm, $\Sim$ is the simulator, and $\Ext$ is the knowledge extractor. The IVXV system adopts the Verificatum mix-net, which utilises the NIZK proposed in \cite{TW10} for verifiable shuffle.
%Let $\PK$ be the public key, $(c_1,\ldots, c_n)$ be the original ciphertexts,
%$(c'_1,\ldots, c'_n)$ be the shuffled ciphertexts. Let $\Pi\in S_n$ and $(r_1,\ldots,r_n)$ the ranomness. The NIZK is defined as
%
%$$
%\NIZK \left\{
%\begin{array}{r}
%(\PK,(c_1,\ldots,c_n),(c'_1,\ldots,c'_n)),(\Pi,(r_1,\ldots,r_n)): \\
%\forall i\in[n]: c'_i= \Rand(\PK,c_{\Pi(i)} ; r_i)
%\end{array}
%\right\}\enspace.
%$$

%\vspace{3pt}
%\noindent\textbf{Universally Composability (UC).}
\subsection{Universally Composability (UC)}
%We provide a review of the UC security framework. The current text is somewhat informal for clarity and brevity. Please refer to \cite{E:Canetti00} for full details.
%! TEX encoding =UTF-8 Unicode
%! TEX root = mainArxiv.tex
%
%
%\subsection{Universally Composable }
%
%
Following Canetti's framework \cite{FOCS:Canetti01}, a protocol is represented as a set of interactive Turing machines (ITMs), each of which represents the program to be run by a participant.
%Protocols that securely carry out a given task are defined in three steps as follows. First, the process of executing a protocol in an adversarial environment is formalized. Next, an ``ideal process'' for carrying out the task at hand is formalized. The parties have access to an ``ideal functionality'' which is essentially an incorruptible ``trusted party'' that is programmed to capture the desired functionality of the task at hand.
%
Let $\exec_{\Pi,\AAA,\ZZZ}$ denote the output of the environment $\ZZZ$ when interacting with parties running the protocol $\Pi$ and real-world adversary $\AAA$. Let $\exec_{\FFF,\SSS,\ZZZ}$ denote output of $\ZZZ$ when running protocol $\phi$ interacting with the ideal functionality $\FFF$ and the ideal adversary $\SSS$.

\begin{definition}
  We say that a protocol $\Pi$ UC-realizes $\FFF$ if for any PPT adversary $\AAA$ there exists an PPT adversary $\SSS$ such that for any environment $\ZZZ$ that obeys the rules of interaction for UC security we have $\exec_{\Pi,\AAA,\ZZZ}\approx \exec_{\FFF,\SSS,\ZZZ}$.
\end{definition}

%\vspace{-10pt}
\section{Security Model}\label{sec:securitymodel}
%\vspace{-5pt}
%! TEX encoding =UTF-8 Unicode
%! TEX root = mainArxiv.tex

Let $n,k\in\NN$ be $\poly(\secp)$. The entities involved in IVXV are a set of voters $\VVV:=\{\Voter_1,\cdots,\Voter_n\}$, a set of trustees $\TTT:=\{\Trustee_1,\cdots,\Trustee_k\}$, the election authority $\EA$, and the auditor $\AU$.
We consider the security of IVXV in the UC framework with static corruption. The security is based on the indistinguishability between real/hybrid world executions and ideal world executions, i.e., for any possible PPT real/hybrid world adversary $\AAA$ we will construct an ideal world PPT simulator $\SSS$ that can present an indistinguishable view to the environment $\ZZZ$ operating the protocol.
%

%! TEX encoding =UTF-8 Unicode
%! TEX root = mainArxiv.tex

\myfullbox{\small Functionality $\Fvote$}{white!40}{white!10}{
\small%
%\footnotesize
\medskip
%The functionality interacts with a set of voters $\VVV:=\{\Voter_1,\ldots,\Voter_n\}$, election authority $\EA$, a set of trustees $\TTT:=\{\Trustee_1,\ldots, \Trustee_k\}$, auditor $\AU$, and the adversary $\SSS$.
It is parameterized with variables $\status$, $\Record$, $\Ballots$, $\tau$, $\JJJ_1$ and $\JJJ_2$. Let $\PPP_{\cor}$ be the set of corrupted parties.

Initially, set $\status=0$, $\Alert:=(0,\ldots, 0)$, $\Record: = \Ballots: = \tau:= \JJJ_1 := \JJJ_2 := \emptyset$.

\medskip
\textbf{Preparation:}
\begin{itemize}
\item Upon receiving input $(\Start,sid)$ from the trustee $\Trustee_j \in \TTT$, set $\JJJ_1 := \JJJ_1 \cup \{\Trustee_j \}$, and send a notification message $(\Start, sid, \Trustee_j )$ to the adversary $\SSS$.

\item Upon receiving $(\Begin, sid)$ from the $\EA$,  if $|\JJJ_1|< k$ ignore the input. Otherwise, send notification message $(\Begin, sid)$ to $\SSS$ and set $\status := 1$.

\end{itemize}

\textbf{Voting:}

\begin{itemize}
\item Upon receiving input $(\Vote,sid, x_i  )$ from a voter $\Voter_i \in \VVV$, if $\status = 1$, set $\Record[i]:= x_i$ and set $\Alert[i] := 1$. Send  a notification message $(\Vote, sid, \Voter_i)$ to the adversary $\SSS$; if $|\TTT\cap\PPP_\cor| \geq t$, send addition message $(\Leak,sid, \Voter_i, x_i)$ to $\SSS$.

Upon receiving $(\Corrupt, sid, \Voter_i)$ from $\SSS$, send $(\Leak,sid, \Voter_i, x_i)$ to $\SSS$. Upon receiving $(\Proceed,sid, \Voter_i, x_i^*)$ from $\SSS$, if $x^*_i = \Record[i]$,
set $\Ballots[i] := \Record[i]$ and $\Alert[i]: = 0$; otherwise, if set $\Ballots[i] := x^*_i$.

\item Upon receiving $(\End, sid)$ from the $\EA$, send notification message $(\End, sid)$ to $\SSS$ and set $\status := 2$.
\end{itemize}

\textbf{Tally:}

\begin{itemize}

\item Upon receiving input $(\Tally,sid)$ from the trustee $\Trustee_j \in \TTT$, if $\status = 2$, set $\JJJ_2 := \JJJ_2 \cup \{\Trustee_i \}$, and send a notification message $(\Tally, sid, \Trustee_j )$ to the adversary $\SSS$.

If $|\JJJ_2| \geq t $, set $\tau\leftarrow \TallyAlg(\Ballots)$.
%\item If $|\JJJ_2 \cap \PPP_{\cor}| + |\PPP_{\cor} | \geq t$:
\begin{itemize}
\item If $\EA\in\PPP_\cor$, send message $(\Leak, sid, \Ballots)$ to the adversary $\SSS$.
\item If $\EA\not\in\PPP_\cor$,
sort entries in $\Ballots$ lexicographically to $\Ballots^*$ and send message $(\Leak, sid, \Ballots^*)$ to the adversary $\SSS$.
\end{itemize}

\item Upon receiving input $(\Result,sid)$ from any party $p$,   if $\tau = \emptyset$ ignore the input. Otherwise, return public delayed output $(\Result, sid, \tau)$ to the requestor.

\end{itemize}

\textbf{Audit:}

\begin{itemize}
\item Upon receiving input $(\Audit,sid)$ from $\AU$, if $\status<2$ ignore the request; otherwise, it sends notification $(\Audit,sid)$ to $\SSS$ and then returns $(\Audit, sid, \valid)$ to $\AU$ if $\HW(\Alert) \leq \delta$; %O(\log \lambda)$; 
else,  it returns  $(\Audit, sid, \invalid)$ to $\AU$, where $\HW$ is the hamming weight function.

\end{itemize}
}{Functionality $\Fvote$\label{fig:Fvote}}
%\vspace{-5pt}

%\vspace{3pt}
%\noindent\textbf{The idea world execution.}

%\subsubsection{The idea world execution.}
\medskip
\noindent\textbf{The idea world execution.}
In the ideal world, the election authority $\EA$,  the voters $\VVV$, the trustee $\TTT$ and the auditor $\AU$ only communicate to an ideal functionality $\Fvote$ during the execution, where $\delta \in \NN$ is an adjustable threshold. The ideal functionality $\Fvote$ accepts a number of commands from $\EA$, $\VVV$, $\TTT$, and $\AU$. At the same time, it informs the adversary $\SSS$ of certain actions that take place and also is influenced by $\SSS$ to elicit certain actions.
As depicted in Fig.~\ref{fig:Fvote}, the ideal functionality $\Fvote$ consists of four phases which are preparation, voting, tally, and audit.
In the preparation phase, the trustees $\TTT:=\{\Trustee_1,\ldots, \Trustee_k\}$ send $(\Start,sid)$ to $\Fvote$ to indicate their presence. This is used to model their key generation participation in practice. To start an election, the $\EA$  sends the command $(\Begin,\sid)$ to $\Fvote$. Note that the election will not start until all the trustees have participated in the preparation.

In the voting phase, the voter $\Voter_i$ sends $(\Vote, \sid, x_i)$ to $\Fvote$. However, this action may be blocked or tampered with by the adversary $\AAA$; in both cases, a flag $\Alert[i]$ is set to $1$. It maintains 2 arrays -- $\Record, \Ballots$. $\Record$ is only used for temporary storage to enable adversarial modification, and the final effective ballots are stored in $\Ballots$.  When the ballot is received from $\Voter_i$, it is temporarily saved in $\Record[i]$; the functionality then asks the adversary $\SSS$ if she wants to proceed, modify, or block it. If $\SSS$ does not reply to the functionality, then the ballot is blocked, and it will not be copied to $\Ballots[i]$  and not counted. Note $\Alert[i]$ is initially set to $1$, and it is only set to $0$ when $\Ballots[i] := \Record[i]$, i.e., the ballots are proceeded without tampering.
 If more than $t$ trustees $\Trustee_j\in\TTT$ are corrupted, i.e., $|\TTT\cap\PPP_\cor| \geq t$, $\Fvote$ directly leaks the voter's choice $x_i$ to $\AAA$. This is used to model the fact that the corrupted trustees can reconstruct the secret key and decrypt all the submitted ballots in practice.
Alternatively, $\AAA$ can send a $\Corrupt$ command to $\Fvote$ to learn the voter's choice $x_i$. This is used to model the fact that if the voter $\Voter_i$'s $\VSD$ is compromised, then his choice may be leaked.
Note that this functionality allows the voters to re-vote an arbitrary number of times before the end of the election.
The $\EA$ can end the election by sending $(\End,sid)$ command to $\Fvote$.

In the tally phase, the trustee $\Trustee_j \in \TTT$ can participate the tally by sending $(\Tally,sid)$ to $\Fvote$. When at least $t$ trustees agree to tally, $\Fvote$  computes the election result $\tau\leftarrow \TallyAlg(\Ballots)$ by invoking the tally algorithm $\TallyAlg(\cdot)$\footnote{The tally algorithm $\TallyAlg$ varies among different elections and countries.}.
If the $\EA$ is corrupted, $\Fvote$ leaks all the ballots $\Ballots$ to $\SSS$, i.e., no voter privacy when the $\EA$ is corrupted. When the $\EA$ is honest,
the adversary only learns lexicographically sorted ballots $\Ballots^*$, which is used to model the privacy level provided by the mix-net.
When $\tau$ is computed, any party can query the election result by sending $(\Result, sid)$ to $\Fvote$.
In the audit phase, the $\AU$ can send the command $(\Audit,sid)$ to $\Fvote$ for auditing. $\HW(\cdot)$ denotes the hemming weight function, and $\HW(\Alert)=\delta$ stands for at most $\delta$ ballots can be tampered or blocked without detection. Otherwise, $\Fvote$ returns $(\Audit,\sid,\invalid)$ to the $\AU$.
This modeling ensures that the difference between the announced tally and the true tally is bounded by small $\delta$ w.r.t. the total number of votes. This is used to model a small tally deviation that is not sufficient to swing the election results in practice. For an election with several billion voters, the deviation could remain as low as hundreds.

\smallskip
\noindent\textbf{Remark.} The above security model implies the following e-voting privacy and verifiability properties. Note that the receipt-freeness property is not modeled above for readability. We refer interested readers to \cite{C:AOZZ15} for a UC model for receipt-freeness. Regarding privacy, in general, the protocol achieves the standard mix-net type of voter anonymity, where a voter's ballot is hidden among all the other cast ballots. When the $\EA$ is compromised, or more than $t-1$ trustees are compromised, voter privacy is no longer guaranteed. Otherwise, if the $\VSD$ of a voter is compromised, then this voter's ballot is leaked to the adversary. In terms of verifiability, assuming the $\ASD$ and $\AU$ cannot be compromised, the protocol achieves end-to-end verifiability against malicious $\EA$ as well as $\VSD$'s in the sense that the announced tally can only deviate $O(\log \lambda)$ votes from the true tally.

%{3pt}
\smallskip
%\subsubsection{The real world execution.}
\noindent{\textbf{The real world execution.}}
The real/hybrid world IVXV protocol $\Pivote$ utilizes a number of supporting components. Those supporting components are modeled as ideal functionalities, and later we will discuss how they are realized in practice in Sec.~\ref{sec:implementation}.
 Similar as the most End-to-End verifiable voting schemes, the IVXV protocol requires a bulletin board functionality $\Fbb$. We distinguish the voters (human) from their voting and auditing devices. Let $(\Fvsd^{\Voter_1}, \ldots, \Fvsd^{\Voter_n})$ and $\Fasd$  denote the \emph{voter supporting device} functionalities associated with the voter $\Voter_i$, $i\in[n]$ and \emph{audit supporting device} functionality, respectively.
In addition, we also abstract the threshold key generation and decryption process as $\Fkeygen$ and $\Fdec$. (Note that we intentionally use $\Fkeygen$ and $\Fdec$ instead of a threshold public-key encryption functionality.  This is mainly because, in Estonian e-voting, the threshold key generation and decryption processes are performed as ceremonies that involve human participates with different assumptions. Our modeling approach is similar to \cite{E:SzePre15}.)
We use the certificate functionality $\mFcert$ to model the Estonian PKI infrastructure. Finally, we introduce the voter emulator functionality $\Fve^{\DDD}$ to capture the voter behavior in an e-voting ceremony.

%\subsubsection{Certificate functionality.}

\medskip
\noindent{\textbf{Certificate functionality.}}
We adopt a multi-session version of the certificate functionality as modeled in \cite{E:Canetti03}. %As depicted in Fig.~\ref{fig:Fcert}, the
The multi-session certificate functionality $\mFcert$ can provide a direct binding between a signature of a message and the identity of the corresponding signer. This corresponds to providing signatures accompanied by ``certificates'' that bind the verification process to the signers' identities. For completeness, we recap $\mFcert$ in Fig.~\ref{fig:Fcert} and refer interested readers to  \cite{E:Canetti03}.

%\vspace{-1.2mm}
\mybox{\small Functionality $\mFcert^{}$}{white!40}{white!10}{
\footnotesize
%\small
\medskip
%The functionality $\mFcert$
It interacts with the VSD functionalities $(\Fvsd^{\Voter_1}, \ldots, \Fvsd^{\Voter_n})$,  the ASD functionality $\Fasd$, the election authority $\EA$, the auditor $\AU$, and the adversary $\AAA$.

\begin{itemize}
\item %{\bf Signature Generation:}
Upon receiving $(\textsc{Sign}, sid, ssid, m)$ from $\Fvsd^{\Voter_i}$, $i\in[n]$, verify that $ssid=(\Voter_i,ssid')$ for some $ssid'$. If not, ignore the request. Otherwise, send $(\textsc{SignNotify}, sid, ssid, m)$ to the adversary $\AAA$. Upon receiving $(\textsc{Signature}, sid, ssid, m,\sigma)$ from $\AAA$, verify that no entry $(ssid, m,\sigma,0)$ is recorded. If it is, then return $(\textsc{Error}, sid)$ to $\Fvsd^{\Voter_i}$ and halt. Else, return $(\textsc{Signature}, sid, ssid, m,\sigma)$ to $\Fvsd^{\Voter_i}$, and record the entry $(ssid, m,\sigma,1)$.

\item %{\bf Signature Verification:}
Upon receiving $(\textsc{Verify}, sid, ssid, m, \sigma)$ from $p \in \{\Fasd, \EA, \AU\}$,
send $(\textsc{VerifyNotify}, sid, ssid, m)$ to the adversary $\AAA$. Upon receiving $(\textsc{Verified}, sid, ssid, m,b^*)$ from $\AAA$, do:

 \begin{itemize}
 \item If $(ssid, m,\sigma,1)$ is recorded then set $b=1$.
 \item Else, if the signer of subsession $ssid$ is not corrupted, and no entry $(ssid, m,\cdot, 1)$ is recorded, then set $b=0$ and record the entry $(ssid, m,\sigma,0)$.
 \item Else, if there is an entry $(ssid, m,\sigma,b')$ recorded, then set $b:=b'$.
 \item Else, set $b:=b^*$, and record the entry $(ssid, m,\sigma,b^*)$.
  \end{itemize}
Return $(\textsc{Verified}, sid, ssid, m,b)$ to $p$.

\end{itemize}

}{The multi-session  functionality for certificate. \label{fig:Fcert}}

%\vspace{-1.2mm}
\mybox{\small Functionality $\Fbb$}{white!40}{white!10}{
\footnotesize
%\small
\medskip
The functionality interacts with the election authority $\EA$, auditor $\AU$, the functionalities $\Fvsd$, $\Fasd$, $\Fdec$, the adversary $\AAA$ and the set of all the other parties $\PPP$. It is parameterized with variables $\Pub$ and $\Priv$.

Initially, set $\Pub = \emptyset$ and $\Priv = \emptyset$.

\begin{itemize}
\item Upon receiving $(\PubPost, sid, m)$ from the $\EA$ or $\Fdec$,  set $\Pub := \Pub \cup \{m\}$ and send notification $(\PubPost, sid, m)$ to the adversary $\AAA$.
\item Upon receiving $(\PrivPost, sid, m)$ from the $\EA$, then set $\Priv := \Priv \cup \{m\}$.

\item Upon receiving $(\Read, sid)$ from a party $p\in\PPP\cup\{\EA, \Fvsd, \Fdec \}$ or $\AAA$, then return  $(\Read, sid, \Pub)$ to the requestor.
\item Upon receiving $(\Read, sid)$ from the $\AU$ or $\Fasd$ or $\Fdec$, then return  $(\Read, sid, (\Pub,\Priv))$ to the requestor.

\end{itemize}

\vspace{8pt}
}{Functionality $\Fbb$ \label{fig:Fbb}}
%\vspace{-15mm}

\smallskip
\noindent{\textbf{BB functionality.}}
%\subsubsection{BB functionality.}
The global \BB functionality $\Fbb$ is presented in Fig.~\ref{fig:Fbb} and consists of private \BB (i.e., $\Priv$) and public \BB (i.e., $\Pub$). In practice, the functionality of the $\Pub$ can be efficiently realized from a conventional \BB functionality using encryption and signature schemes.  The introduction of $\Priv$ allows IVXV to enable verifiability while still achieving coercion resistance. Namely, only the honest party can access $\Priv$. In this work, we only allow the $\EA$ to post a message on the $\Fbb$. Meanwhile, any party can read the messages posted on $\Pub$ of $\Fbb$, while only $\AU$ and $\Fasd$ can access  $\Priv$. % of $\Fbb$.

%For readability, we only allow the $\EA$ to post a message on the $\Fbb$. Note that this BB functionality can be efficiently realized from a conventional BB functionality using encryption and signature schemes. The functionality $\Fbb$ is presented in Fig.~\ref{fig:Fbb}, any party can read the messages posted on $\Pub$ of $\Fbb$, while only $\AU$ and $\Fasd$ can access  $\Priv$ of $\Fbb$.

\begin{table*}[t!]%[ht]
\begin{center}
\begin{threeparttable}
%\vspace{-0.6mm}
%\vspace{10pt}
\caption{Voter Behaviour Statistics$^{*}$}\label{table:votedis}
%\resizebox{\textwidth}{15mm}{
{
%\footnotesize
\begin{tabular}{l @{\quad}l @{\quad}l @{\quad}l||l @{\quad}l @{\quad}l @{\quad}l}
\hline\rule{0pt}{12pt} %\toprule
Pattern & KOV2013 & EP2014 & RK2015&Pattern & KOV2013 & EP2014 & RK2015\\
\hline \hline\rule{0pt}{12pt} %\midrule
%%%%%%%%%%%%%%%%%%%%%%%%%%%%%

V& {94.9816\%}& {94.6108\%}& { 93.8592}\% & VVVV& 0.0090\%& 0.0107\%& 0.0119\% \\
VC& 3.0858\%& 3.6994\%& 3.8563\% & VCVCVC& 0.0082\%& 0.0078\%& 0.0091\%\\
VV& 1.4977\%& 1.1885\%& 1.7117\% & VVCVC& 0.0037\%& 0.0068\%& 0.0062\%\\
VVC& 0.1853\%& 0.2482\%& 0.2941\%& VVVVC& 0.0022\%& 0.0048\%& 0.0006\%\\
VCVC& 0.0523\%& 0.0795\%& 0.0686\% & VCVVC& 0.0015\%& 0.0029\%& 0.0023\% \\
VVV& 0.1069\%& 0.0659\%& 0.1031\%& VVVVV& 0.0045\%& 0.0019\%& 0.0017\%\\
VCV& 0.0277\%& 0.0359\%& 0.0340\%& {VCVCVCVC}& 0.0007\%& 0.0019\%& 0.0011\%\\
VVVC& 0.0135\%& 0.0107\%& 0.0164\%& VVVCV& 0.0007\%& 0.0019\%& 0.0000\%\\
\hline
%\end{longtable}
\end{tabular}
%}
}
\vspace{1ex}

\centering
\begin{tablenotes}
  \footnotesize%\small      
\begin{minipage}{0.4\textwidth}%{2in}
	%\begin{itemize}
	\item[-] KOV: local municipal elections;
\item [-] EP: European Parliament;
\item [-] RK: Estonian Parliamentary elections.
	%\end{itemize}
  \end{minipage}%
%\hspace*{17mm}
  \begin{minipage}{0.4\textwidth}%{2in}
 	%\begin{itemize}
 \item [-] C: check;
	\item[-] V: vote;
	%\end{itemize}
\end{minipage}
\vspace{1ex}
\item[*] We remark that there are various kinds of voters' behaviour, we just list some voters' behaviour with significant probability in Table~\ref{table:votedis}, the rest are insignificant  and we omit these behaviours' probability.
\end{tablenotes}
\end{threeparttable}

\end{center}

\vspace{-10pt}
\end{table*}
%\vspace{-5pt}

%\vspace{-1em}
\medskip
\noindent{\bf Voter emulator functionality.}
%\subsubsection{Voter emulator functionality.}
The voter emulator functionality $\Fve^\DDD$  samples voter behavior according to a distribution $\DDD$. Table~\ref{table:votedis} depicts the voter behaviour statistics collected from recent Estonian national-wide elections. The voter has two types of actions, (i) `V', which stands for voting, and (ii) `C', which stands for checking. For instance, in the RK2015 election, $93.8592\%$ voters have action pattern `V', which means that those voters voted their ballots without checking; whereas $3.8563\%$ voters have action pattern `VC', which means that those voters voted their ballots and checked; similarly, `VVC' means that the voters voted once without checking, and he then voted the second time and checked the ballot.

Notably, as a coercion prevention measure, the voter is allowed to re-vote as many times as he wants. After every vote casting, he can check the last vote $0$ or more times.
For the sake of probabilistic analysis, the string of script $\Act_i$ is \delete{be} expressed as the regular expression $V(V|C)^*$.\footnote{In practice, the maximal number of verifications per vote is limited to $3$, but we ignore this limitation here for notation simplicity.}
%

%The decision concerning success or failure of verification is taken in the head of the voter, and the machine does not know the outcome of this decision. The voter can stop the process after every step of vote casting or verification.

\mybox{\small Functionality $\Fvsd^{\Voter_i}$}{white!40}{white!10}{
%\footnotesize
\small
\medskip

The functionality $\Fvsd^{\Voter_i}$ interacts with the voter $\Voter_i\in\VVV$, the $\EA$, and the adversary $\AAA$. It is parameterized with a PPT Turing machine $\TMH$ and an internal state $\st$.
Initially, set $\st:= \emptyset$.
\begin{itemize}
%\item Upon receiving $(\Init, sid, \TM, \st)$ from the $\EA$, store $\TMH$ and $\st$, and then set $\status:=1$.
\item Upon receiving $(\SignEnc, sid , x)$ from the voter $\Voter_i$, send notification $(\SignEnc, sid, \Voter_i)$ to the adversary $\AAA$.
Upon receiving $(\Corrupt, sid)$ from $\AAA$, send $(\Leak, sid, \Voter_i, x)$ to $\AAA$. Upon receiving $(\Tamper, sid, \TM)$ from $\AAA$,
if $\TM=\emptyset$, compute $(y, z, \st')\leftarrow\TMH(\Voter_i, \st, x)$; Otherwise, send  compute $(y, z,  \st')\leftarrow\TM(\Voter_i,\st, x)$.
 It updates $\st:=\st'$; sends $(\Ballot, sid, y)$ to the $\EA$; and sends $(\Receipt ,sid, z)$ to the voter $\Voter_i$.

\end{itemize}

}{Functionality $\Fvsd^{\Voter_i}$ \label{fig:Fvsd}}

%\vspace{-5pt}
\smallskip
%\subsubsection{$\VSD$ functionality.}

\noindent{\bf   $\VSD$ functionality.}
Each voter has its own unique $\VSD$, as the Estonian national ID cards are modeled as a part of the $\VSD$.  The voter $\Voter_i$'s  $\VSD$ is modeled as the functionality $\Fvsd^{\Voter_i}$, as depicted in Fig.~\ref{fig:Fvsd}. $\Fvsd^{\Voter_i}$ is parameterized with a PPT Turning machine $\TMH$, which is used to model the software running on the $\VSD$. The functionality of $\TMH$ is encrypting and signing the ballots, and the details are provided later. Note that $\Fvsd^{\Voter_i}$ can be corrupted. When it is corrupted, the voter's choice is leaked to the adversary $\AAA$. Moreover, the adversary $\AAA$ is able to replace $\TMH$ with any arbitrary Turning machine $\TM$, and $\Fvsd^{\Voter_i}$ will use $\TM$ for the execution. The state $\st$ is introduced to model stateful algorithms. After execution, $\Fvsd^{\Voter_i}$ sends the voter $\Voter_i$ the receipt by the command $(\Receipt ,sid, z)$ and sends $\EA$ the ballot using the command $(\Ballot, sid, y)$.

\mybox{\small Functionality $\Fve^{\DDD}$}{white!40}{white!10}{
\footnotesize
%\small
\medskip
The functionality interacts with voters $\{\Voter_1,\ldots,\Voter_n\}$ and the adversary $\AAA$. It is parameterized with a voter action distribution $\DDD$.
\begin{itemize}
\item Upon receiving  $(\Emulate, sid)$ from a voter $\Voter_i\in\VVV$, sample $\Act_i\leftarrow\DDD$ and  return  $(\Emulate, sid, \Act_i)$ to $\Voter_i$.
\end{itemize}
\vspace{8pt}
}{Functionality $\Fve^{\DDD}$ \label{fig:Fve}}

\begin{figure}[ht] % ’ht’ tells LaTeX to place the figure ’here’ or at the top of the page
\vspace{-2mm}
\centering % centers the figure
%\scriptsize
\small
\begin{tikzpicture}
    \node[state, initial] (q0) {Begin};
    \node[state, accepting, right of = q0, xshift=1.5cm] (q1) {Voted};
    \node[state, accepting, right of = q1, xshift=1.5cm] (q2) {Checked};

    \draw   (q0) edge[above] node{$V$} (q1)
            (q1) edge[loop above] node{$V$} (q1)
            (q1) edge[bend left, above] node{$C$} (q2)
            (q2) edge[bend left, above] node{$V$} (q1)
            (q2) edge[loop above] node{$C$} (q2);
\end{tikzpicture}
\caption{FSM for voting and checking}
\label{fig:FSM}
\end{figure}
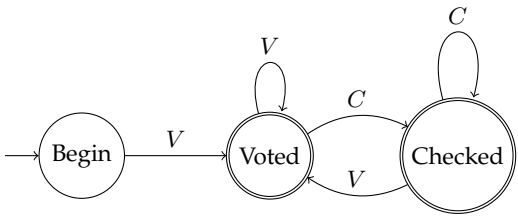

\smallskip
\noindent\textbf{Modelling Voters (human).}\label{sec:human}
%\subsubsection{Modelling Voters (human).}\label{sec:human}
%! TEX encoding =UTF-8 Unicode
%! TEX root = mainArxiv.tex
%\documentclass{article}
%
%\newcommand{\SignEnc}{\textsc{SignEnc}}
%\newcommand{\Receipt}{\textsc{Receipt}}
%\newcommand{\Audit}{\textsc{Audit}}
%\newcommand{\Fvsd}{\mathcal{F}_{\textsc{VSD}}}
%\newcommand{\Fasd}{\mathcal{F}_{\textsc{ASD}}}
%
%\usepackage{tikz}
%\usetikzlibrary{automata, positioning, arrows}
%
%\tikzset{->, initial text=$ $}
%
%\begin{document}
We assume that the voters as human beings are not capable of performing complicated cryptographic operations or generating high entropy randomness.
In fact, in IVXV,  the voters are only required to perform string forwarding and matching.
We model the voter (human) as a finite state machine (FSM),  \fbox{$\VOTE(\Act_i,sid, x)$} by taking as input the script $\Act_i$, session id $\sid$, and the choice $x$ of the voter, where the $\Act_i$  is an output of $\Fve^{\DDD}$ (see Fig.~\ref{fig:Fve}.)
The machine FSM then starts executing $\Act_i$ character by character, and the logic of implementing the voting and verification via the script $\Act_i$ is depicted in Fig.~\ref{fig:FSM}.
The action commands corresponding to the characters are interpreted as follows.
\begin{itemize}
    \item $V$ (\textbf{Vote}): Send message $(\SignEnc, sid , x)$  to functionality $\Fvsd$, and receive $(\Receipt ,sid, z)$ from $\Fvsd$.
    \item $C$ (\textbf{Check}): Send message $(\Audit, sid, z, x)$ to functionality $\Fasd$, and obtain $(\Audited, sid, x^*)$ from $\Fasd$. If $x^*= x$ continue; otherwise, sends $\Complain$ to the auditor $\AU$ and halt.
\end{itemize}

The decision concerning the success or failure of verification is taken in the head of the voter, and the machine does not know the outcome of this decision. The voter can stop the process after every step of vote casting or verification.

%\end{document}

%\subsubsection{$\TMH$ description.}
\smallskip
\noindent\textbf{$\TMH$ description.}
As depicted in Fig.~\ref{fig:TM}, the Turing machine $\TMH$ processes the $\Fvsd^{\Voter_i}$ requirements from the voter $\Voter_i$ as follows.
The Turing machine $\TMH$ first fetch the public key $\PK$ from $\Fbb$. It samples a random $r\gets\Zq$, then encrypts and outputs $c\gets\Enc(\PK,x;r)$ for the voter choice $x$. Once obtains the ciphertext $c$, the Turing machine $\TMH$ sends the message $(\Sign, sid ,ssid, c)$ to functionality $\mFcert$ and receives the feedback $(\Signature, \linebreak sid, ssid, c,\sigma)$ from $\mFcert$. Lastly, the Turing machine $\TMH$ will output the triple $(y,z,\emptyset)$, where $y$ is the ciphertext and signature triple $(\ssid,c,\sigma)$, the receipt $z$ is the ciphertext and random triple $(\ssid, c, r)$. The initial $\emptyset$ implies $\TMH$ is stateless.

%Please refer to Fig.~\ref{fig:TM} for more details.
%\vspace{-1.2mm}
\mybox{\small $\TMH(\Voter_i, \st,x )$  }{white!40}{white!10}{
\footnotesize
%\small
\medskip
\vspace{-8pt}
%\textbf{Vote:}
\begin{itemize}
\item Send  $(\Read, sid)$ to functionality $\Fbb$, and obtains $(\Read,sid, \Pub)$ from $\Fbb$. It fetches $\PK$ from $\Pub$;
\item Pick a random $r\leftarrow \ZZ_p$, and encrypt $x$ as $c \leftarrow\Enc_{\PK}(x;r)$;
\item  Send message $(\Sign, sid ,ssid, c)$  to functionality $\mFcert$, and receive $(\Signature ,sid,ssid, c,\sigma)$ from $\mFcert$, where $ssid =
(\Voter_i , ssid' )$ for some $ssid'$;
\item Set $y:= (ssid, c,\sigma)$ and $z:= (ssid, c, r)$;
\item Return $(y,z,\emptyset)$.
\end{itemize}

}{ $\TMH$  description \label{fig:TM}}

\smallskip
\noindent\textbf{Remark.} Although it is controversial whether it is a good idea to let an ideal functionality directly interact with another ideal functionality in the UC framework, technically, such type of modeling is not wrong and can be found in many UC modeling papers, e.g. \cite{C:BMTZ17,EC:BGMTZ18}.

%
%\mybox{\small $\Act$ script}{white!40}{white!10}{
%\scriptsize
%\textbf{Vote:}
%\begin{itemize}
%\item  Send message $(\SignEnc, sid , x)$  to functionality $\Fvsd$, and receive $(\Receipt ,sid, z)$ from $\Fvsd$.
%\end{itemize}
%
%\textbf{Check:}
%\begin{itemize}
%\item Send message $(\Audit, sid, z, x)$ to functionality $\Fasd$.
%
%\end{itemize}
%
%}{$\Act$ script \label{fig:Act}}

%\medskip
\vspace{25pt}
%\subsubsection{$\ASD$ functionality.}
\noindent\textbf{$\ASD$ functionality.}
Although each voter may have its own $\ASD$, the corresponding operation is universal. Hence we use a single functionality $\Fasd$ to model all the $\ASD$s. $\Fasd$ is always trusted in our setting.
As depicted in Fig. \ref{fig:Fasd}, the voter $\Voter_i\in \VVV$ can send the command $(\Audit, sid , z)$ to $\Fasd$ for audit, and $\Fasd$ parses the receipt $z$ as $(\ssid,c,r)$ and then verify that $\ssid=(\Voter_i,\ssid')$ for some $\ssid'$. Subsequently, the functionality $\Fasd$ fetches the corresponding record from $\Fbb$, and verify the signature by sending  $(\Verify, sid, ssid, m, \sigma)$ to $\mFcert$. Upon success,  $\Fasd$ computes $m \leftarrow \TDec(\PK, c,r)$ and returns $(\Audited, \linebreak sid, m)$ to $\Voter_i$.

%\vspace{-1.2mm}
\mybox{Functionality $\Fasd$}{white!40}{white!10}{
%\footnotesize
\small
\medskip
The functionality interacts with a set of voters $\VVV:=\{\Voter_1,\ldots,\Voter_n\}$, the functionality $\Fbb$, and the adversary $\AAA$. It is parameterized with an algorithm $\TDec$ (cf. Sec.~\ref{sec:preliminaries}).

\begin{itemize}
\item Upon receiving $(\Audit, sid, z)$ from a voter $\Voter_i\in\VVV$, parse $z$ as $(ssid, c, r)$, and verify that $ssid = (\Voter_i, ssid')$ for some $ssid'$. If not, return $(\Audited, sid, \textsc{Fail})$ to $\Voter_i$ and halt. Send $(\Read, sid)$
to functionality $\Fbb$, and obtains $(\Read,sid, (\Pub,\Priv))$ from $\Fbb$. It fetches $(ssid, c,\sigma)$ from $\Priv$ and $\PK$ from $\Pub$; if this step fails, return $(\Audited, sid, \textsc{Fail})$ to $\Voter_i$; otherwise, send $(\textsc{Verify}, sid, ssid, m, \sigma)$ to functionality $\mFcert$ and receive $(\textsc{Verified}, sid, ssid, m,b)$ from $\mFcert$.

If $b = 0$,  return $(\Audited, sid, \textsc{Fail})$ to $\Voter_i$; else, compute $m \leftarrow \TDec(\PK, c,r)$.
Return $(\Audited, sid, m)$ to $\Voter_i$.

\end{itemize}

}{Functionality $\Fasd$\label{fig:Fasd}}

%The functionality then requests the adversary $\AAA$ to produce the ballot by forwarding the command $(\SignEnc, sid , x)$. The adversary $\AAA$ after that determines whether to corrupt the desktop software $\TM$, and returns $\Fvsd$ the decision. In this setting, if $\AAA$ determines to corrupt $\TM$, then the functionality $\Fvsd$ leaks the information to $\AAA$ by the command $(\Leak, sid, \Voter_i, x)$, meanwhile, $\Fvsd$ obtains the encrypted ballot $y$ along with the receipt $z$ and state $\st'$ by computing $(y, z,  \st')\leftarrow\TM(\st, x)$. Otherwise, $\Fvsd$ directly compute $(y, z,  \st')\leftarrow\TMH(\st, x)$, please refer to Fig.\ref{fig:TM} for more details about $\TMH$.

%\medskip
\vspace{15pt}

%\subsubsection{Threshold key generation functionality.}
\noindent\textbf{Threshold key generation functionality.}
$\Fkeygen$ is aimed to provide the public key $\PK$ for voters and provide the share of secret key $\SK_j$ ($j\in [k]$) via secret sharing for trustees, as depicted in Fig.~\ref{fig:Fkeygen}.
The trustee $\Trustee_j\in\TTT$ can send the command $(\Ready, sid)$ to $\Fkeygen$ for key generation. Once all the $k$ trustees have participated the key generation, $\Fkeygen$  generates $(\PK,\SK)\leftarrow \KeyGen(1^\lambda)$ and computes $(\SK_1,\ldots, \SK_k)\leftarrow\Deal(\SK)$. Lastly, $\Fkeygen$ returns trustee $\Trustee_j$ message $(\PrivKey,sid, \SK_j)$ for $j\in[k]$. Furthermore, the $\EA$ can send the command $(\PubKey,\sid)$ to the $\Fkeygen$ for the public key $\PK$.

\mybox{Functionality $\Fkeygen$}{white!40}{white!10}{
%\footnotesize
\small
\medskip

The functionality interacts with trustees $\{\Trustee_1,\ldots, \Trustee_k\}$, the $\EA$, and the adversary $\AAA$.
It is parameterised with variable $\JJJ$ and algorithms $\KeyGen()$ and $\Deal()$.

Initially, set $\JJJ:=\emptyset$.

\begin{itemize}
\item Upon receiving $(\Ready, sid)$ from the trustee $\Trustee_j\in\TTT$, set $\JJJ:=\JJJ \cup \{\Trustee_j\}$.

\item If $|\JJJ| = k$, generate $(\PK,\SK)\leftarrow \KeyGen(1^\lambda)$. Compute $(\SK_1,\ldots, \SK_k)\leftarrow\Deal(\SK)$. For $j\in[k]$, send trustee $\Trustee_j$ message $(\PrivKey,sid, \SK_j)$.

\item Upon receiving $(\PubKey,sid)$ from the $\EA$, if $\PK$ is not defined yet, ignore the request. Otherwise, it  sends $(\PubKey,sid, \PK)$ to the $\EA$.

\end{itemize}

}{Functionality $\Fkeygen$\label{fig:Fkeygen}}

%\vspace{-5pt}

\medskip
%\subsubsection{Threshold decryption functionality.}

\noindent\textbf{Audible threshold decryption functionality.}
$\Fdec$ is used to model the audible threshold decryption process, as depicted in Fig.~\ref{fig:Fdec}. The trustee $\Trustee_j\in\TTT$ can send the command  $(\Key, sid, \SK_j )$ to $\Fdec$ to participate the decryption. When more than $t$ trustees participate the description, $\Fdec$  reconstructs $\SK\leftarrow \Reconstruct(\KKK)$.
It then fetches $(c'_1,\ldots,c'_n)$ and $\PK$ from $\Fbb$. After verifying that  $(\PK,\SK) \in \RRR_{\PK}$, for $i\in [n]$, it computes $m_i\leftarrow\Dec(\SK, c'_i)$. The functionality allows  the adversary $\AAA$ to tamper the decryption process; however, it will be detected, as in practice the deception correctness is ensured by NIZK proofs.  Finally, it posts $(\PubPost, sid, (m_1,\ldots, m_n))$ to $\Fbb$.

\mybox{Functionality $\Fdec$}{white!40}{white!10}{
%\footnotesize
\small
\medskip
The functionality interacts with trustees $\{\Trustee_1,\ldots, \Trustee_k\}$, the $\EA$, , the auditor $\AU$,  and the adversary $\AAA$.
It is parameterised with variable $\KKK$, $b$, the decryption algorithm $\Dec()$ and the share reconstruction $\Reconstruct()$.

Initially, set $\KKK:=\emptyset$ and $b=\bot$.

\begin{itemize}
\item Upon receiving $(\Key, sid, \SK_j )$ from the trustee $\Trustee_j\in\TTT$, set $\KKK:=\KKK \cup \{\SK_j\}$.

\item If $|\JJJ| \geq  t$, compute $\SK\leftarrow \Reconstruct(\KKK)$; Send $(\Read, sid)$
to functionality $\Fbb$, and obtains $(\Read, sid, (\Pub,\Priv))$ from $\Fbb$. It fetches $(c'_1,\ldots,c'_n)$ from $\Priv$ and $\PK$ from $\Pub$;  If $(\PK,\SK) \not\in \RRR_{\PK}$ ignore the request. Otherwise, for $i\in[n]$, compute $m_i\leftarrow\Dec(\SK, c'_i)$;
it sends $(\textsc{Dec},sid,(m_1,\ldots, m_n))$ to the adversary $\AAA$.

Upon receiving $(\textsc{Dec},sid,(m^*_1,\ldots, m^*_n))$ from $\AAA$, send $(\PubPost, sid, (m^*_1,\ldots, m^*_n))$ to $\Fbb$. If $\exists i \in[n]$ s.t. $m_i \neq m^*_i$, set $b:=\invalid$; otherwise, set $b:=\valid$.

\item Upon receiving $(\Audit, sid)$ from the auditor $\AU$, return $(\Audit, sid, b)$ to the requestor.

\end{itemize}
%\zlinote{add the definition of $(\PK,\SK)\notin \RRR_{\pk}$}
%\zlinote{Actually, we didn't use the $\Dec$ under $\SK$, and we didn't need to obtain $\SK$ via $\Reconstruct$, so can we ignore the item 2?}
}{Functionality $\Fdec$\label{fig:Fdec}}

%\input{model}
%\vspace{-10pt}
\section{The Estonian IVXV I-voting Scheme}\label{sec:protocol}
%\vspace{-5pt}
% informal description
%! TEX encoding =UTF-8 Unicode
%! TEX root = mainArxiv.tex

On the conceptual level, the Estonian IVXV Internet voting is still mimicking double envelope postal voting as it was in 2005. The inner privacy ensuring envelope is replaced by encryption with the central system's public key, whereas the outer authentication and integrity providing envelope are replaced by the voter's digital signature~\cite{heiberg2016improving,DBLP:conf/voteid/HeibergPW15}. %E:HeiParWil15
However, two major scheme updates have occurred since 2005. First, in 2013, it became possible to individually verify vote integrity in the digital ballot box using an independent mobile device~\cite{HW11}. Secondly, in 2017, server-side transparency was greatly improved by adding several independently auditable features like provable vote decryption and commitments of digital ballot box actions~\cite{heiberg2016improving}.
%
%Before describing the formal Estonian IVXV e-voting scheme, we first highlight core tools with their corresponding functions. Correctly, the core system consists of the Voting Application (VoteApp), the Vote Forwarding Server ($\VFS$), the Vote Storage Server ($\VSS$) and the Tabulation Application ($\TA$) with the Hardware Security Module ($\HSM$) for private key protection.
%
%The online components log to the Log Monitor ($\LOG$), and there is a OCSP responder ($\OCSP$) that provides both certificate validation and time-marking services.
In the following Sec.~\ref{sec:uc evoting}, we will provide the UC description to the Estonian e-voting w.r.t. the documentation \cite{IVXV}. In  Sec.~\ref{sec:implementation}, we show how those ideal functionalities are realized in practice.

\subsection{UC Description for Estonian IVXV Voting scheme}\label{sec:uc evoting}

The Estonian IVXV scheme $\Pivote$ is depicted in Fig.~\ref{fig:main_protocol}. It consists of four phases: preparation, voting, tally, and audit. The entities involved in the $\Pivote$ protocol are the election authority $\EA$, the voters $\VVV$, the trustees $\TTT$, and the auditor $\AU$. Meanwhile, the protocol also uses a number of supporting components modelled as ideal functionalities, i.e. $\Fbb, \mFcert,\{\Fvsd^{\Voter_i}\}_{i\in[n]}, \Fasd, \Fkeygen, \Fdec, \Fve^{\DDD}$, where $\DDD$ is the Estonian voter behaviour statistics as depicted in Table~\ref{table:votedis}.

{

\myfullbox{\small Estonian E-voting Scheme $\Pivote$}{white!40}{white!10}{
%\footnotesize
%\scriptsize
\small
%\medskip

\textbf{Preparation:}

\begin{itemize}
\item Upon receiving $(\Start, sid)$ from the environment $\ZZZ$, the trustee $\Trustee_j \in \TTT$ sends message $(\Ready, sid)$ to functionality $\Fkeygen$.

\item Upon receiving $(\Begin, sid)$ from the environment $\ZZZ$, the $\EA$ creates an empty array $\BBB := \emptyset$. The $\EA$ then sends $(\PubKey, sid)$ to functionality $\Fkeygen$, and it receives $(\PubKey, sid, \PK)$  from $\Fkeygen$. It then sends $(\PubPost, sid, \PK)$ to $\Fbb$. %The $\EA$ then sends $(\Init, sid, \TM, \emptyset)$ to the $\Fvsd$, where $\TMH$ is defined in Fig.~\ref{fig:TM}.

\end{itemize}

\textbf{Voting:}

\begin{itemize}
\item Upon receiving $(\Vote,sid, x)$ from the environment $\ZZZ$, the voter $\Voter_i \in\VVV$
sends message $(\Emulate, sid)$ to functionality $\Fve^{\DDD}$ and obtains $(\Emulate, sid, \Act_i)$ from $\Fve^{\DDD}$.
Execute  \fbox{$\VOTE(\Act_i,sid, x)$} as described in Sec.~\ref{sec:human}, Fig.~\ref{fig:FSM} to interact with $\Fvsd^{\Voter_i}$ and $\Fasd$. %(cf. Fig.~\ref{fig:TM}).

\item Upon receiving $(\Ballot, sid, y)$ from the functionality $\Fvsd^{\Voter_i}$, the $\EA$ parses $y$ as $(ssid, c, \sigma)$ where $ssid = (\Voter_i, ssid')$ for some $i\in[n]$. It then sends $(\Verify, sid, ssid, c, \sigma)$ to functionality $\mFcert$ and obtains $(\textsc{Verified}, sid, ssid, c,b)$ from $\mFcert$. If $b=0$, it halts; otherwise, it sets $\BBB[i]:= c$ and sends $(\PrivPost,sid, (ssid, c,\sigma))$ to $\Fbb$.

\item Upon receiving $(\End,sid)$ from the environment $\ZZZ$, the $\EA$ for $i\in[n]$, defines $c_i:= \BBB[i]$. It then picks a random permutation $\Pi \leftarrow S_n$; for $i\in[n]$, it picks a random $r_i\leftarrow \ZZ_p$ and set the shuffled ciphertext $c'_i := \Rand(\PK,c_{\Pi(i)} ; r_i)
 $. The $\EA$ then generates the corresponding NIZK proof
$$
\footnotesize{
  \pi \leftarrow \NIZK.\Prov \left\{
  \begin{array}{r}
   \PK, \{c_i\}_{i\in[n]}, \{c'_i\}_{i\in[n]}, (\Pi, \{r_i\}_{i\in[n]}):\\
     \forall i\in[n]: \; c'_i = \Rand(\PK,c_{\Pi(i)} ; r_i)
    \end{array}
\right\}\enspace.
}$$

After that, it then sends $(\PrivPost, sid, ((c'_1,\ldots, c'_n), \pi ))$ to $\Fbb$.
\end{itemize}

\textbf{Tally:}

\begin{itemize}

\item Upon receiving $(\Tally,sid)$ from the environment $\ZZZ$, the trustee $\Trustee_j \in\TTT$ sends message $(\Key, sid)$ to functionality $\Fdec$.

%\item Upon receiving $(\Decrypt,sid)$ from the environment $\ZZZ$, the $\EA$ sends $(\Decrypt, sid, \PK, (c'_1,\ldots, c'_n))$
%to the functionality $\Fdec$.

%\item Upon receiving $(\Msg, sid, (m_1,\ldots, m_n) )$ from the functionality $\Fdec$, the $\EA$ computes $\tau\leftarrow \TallyAlg(m_1,\ldots, m_n)$. It then sends $(\PubPost, sid, (\tau, (m_1,\ldots, m_n ) ) )$ to $\Fbb$.

\item Upon receiving $(\Result, sid)$ from the environment $\ZZZ$, any party $p$ sends $(\Read, sid)$
to functionality $\Fbb$, and obtains $(\Read,sid, \Pub)$ from $\Fbb$. $p$ then fetches $(m_1,\ldots, m_n)$ from $\Pub$. It then computes $\tau\leftarrow \TallyAlg(m_1,\ldots, m_n)$ and returns $(\Result, sid, \tau)$ to $\ZZZ$.

\end{itemize}

\textbf{Audit:}

\begin{itemize}

\item Upon receiving $(\Audit, sid)$ from the environment $\ZZZ$, the auditor $\AU$ sends  $(\Read, sid)$
to functionality $\Fbb$, and obtains $(\Read,sid, (\Pub,\Priv))$ from $\Fbb$. It fetches $\PK$ from $\Pub$, and verifies:
\begin{itemize}

\item For all the $(ssid_\ell, c_\ell,\sigma_\ell)$ in $\Priv$, send $(\textsc{Verify}, sid, ssid_\ell, c_\ell, \sigma_\ell)$ to functionality $\mFcert$ and receive $(\textsc{Verified}, sid, ssid_\ell, m_\ell,b_\ell)$ from $\mFcert$. If $b_\ell = 0$, return $(\Audited, sid, \invalid)$ to $\ZZZ$ and halt.

\item For $i\in[n]$, define the last ballot $c_{\ell^*}$ for some $\ell^*$ sent by voter $\Voter_i$ as $c_i$.

\item Check  $\NIZK.\Ver((c_1,\ldots, c_n), (c'_1,\ldots, c'_n), \PK, \pi) = b^*$. If $b^*=0$,  return $(\Audited, sid, \invalid)$ to $\ZZZ$.

\item Send $(\Audit, sid)$ to the functionality $\Fdec$, obtaining  $(\Audit, sid, b)$ from it. If $b=\invalid$, return $(\Audited, sid, \invalid)$ to $\ZZZ$ and halt.

\item If it received $\Complain$ message from any voter $\Voter_i\in\VVV$, return $(\Audited, sid, \invalid)$ to $\ZZZ$ and halt.

\item Otherwise, return  $(\Audited, sid, \valid)$ to $\ZZZ$.

\end{itemize}

\end{itemize}
}{Estonian E-voting Scheme $\Pivote$ \label{fig:main_protocol}}
}

\bigskip

In the preparation phase, the trustee $\Trustee_j \in \TTT$ for initializing the e-voting scheme by sending the command $(\Ready, sid)$ to functionality $\Fkeygen$ for key generation. Upon receiving $(\Begin, sid)$, the $\EA$ fetches the generated public key $\PK$ from $\Fkeygen$, and then it posts $\PK$ to the $\Fbb$.
Meanwhile, the $\EA$ creates $\BBB$, and $\BBB[i]$ will be used to record the last received ballot of voter $\Voter_i\in\VVV$, $i\in[n]$.

In the voting phase, the voter $\Voter_i\in\VVV$ first queries $\Fve^{\DDD}$ to obtains a script $\Act_i$. It runs  \fbox{$\VOTE(\Act_i,sid, x)$} as described in Sec.~\ref{sec:human}, Fig.~\ref{fig:FSM} to interact with $\Fvsd^{\Voter_i}$ and $\Fasd$. Note that if the verification/checking fails w.r.t. the output of $\Fasd$, the voter $\Voter_i$ sends a $\Complain$ message to $\AU$.
The $\Fvsd^{\Voter_i}$ encrypts and signs the ballot, it then sends the ballot to the $\EA$ via $(\Ballot, sid, y)$, where $y=(ssid, c, \sigma)$. The $\EA$ first checks the validity of the signature, by verifying that  $ssid = (\Voter_i, ssid')$ and sending $(\Verify, sid, ssid, c, \sigma)$ to functionality $\mFcert$. The $\EA$ then posts the ballots on the $\Fbb$.

Upon receiving $(\End, sid)$, the $\EA$ picks a random permutation $\Pi \leftarrow S_n$; for $i\in[n]$, it picks a random $r_i\leftarrow \ZZ_p$ and set the shuffled ciphertext $c'_i := \Rand(\PK,c_{\Pi(i)} ; r_i)
 $. The $\EA$ then generates the corresponding NIZK proof
$$
  \pi \leftarrow \NIZK.\Prov \left\{
  \begin{array}{r}
   \PK, \{c_i\}_{i\in[n]}, \{c'_i\}_{i\in[n]}, (\Pi, \{r_i\}_{i\in[n]}): \\
     \forall i\in[n]: \; c'_i = \Rand(\PK,c_{\Pi(i)} ; r_i)
    \end{array}
\right\}\enspace.
$$
After that, it then posts $(c'_1,\ldots, c'_n), \pi )$ to the $\Fbb$.

In the tally phase, the trustee $\Trustee_j \in\TTT$ sends message $(\Key, sid)$ to functionality $\Fdec$ to decrypt the ballots. When more than $t$ trustee participated the decryption, $\Fdec$ decrypts the ballots and post the decrypted shuffled ballots $(m_1,\ldots, m_n)$ on the $\Fbb$. After that, anyone can fetch  $(m_1,\ldots, m_n)$ from $\Fbb$ and then computes the election result $\tau\leftarrow \TallyAlg(m_1,\ldots, m_n)$.

In the audit phase, the auditor $\AU$ fetches all the election transcripts from the $\Fbb$ and it returns $\valid$ if and only if:

\begin{itemize}

\item All the $(ssid_\ell, c_\ell,\sigma_\ell)$ in $\Priv$ have valid signatures.

\item For $i\in[n]$, define the last ballot $c_{\ell^*}$ for some $\ell^*$ sent by voter $\Voter_i$ as $c_i$.

\item Check  $\NIZK.\Ver((c_1,\ldots, c_n), (c'_1,\ldots, c'_n), \PK, \pi) = 1$.

\item No $\Complain$ message received from any voter $\Voter_i\in\VVV$.

\end{itemize}

%implementation
%! TEX encoding =UTF-8 Unicode
%! TEX root = mainArxiv.tex

%\vspace{-10pt}
\subsection{Security Analysis}
%\vspace{-5pt}
We first give an intuition why the IVXV protocol is secure. Let's examine the optimal attacker strategy for IVXV. Note that the adversary also knows the voter behavior distribution as shown in Table~\ref{table:votedis}, where the voter with a probability around 94\% just submits one vote and never does anything else. Apparently, just manipulating the first vote gives the attacker a success probability of about 94\%.
But the adversary knows around 1.5\% of voters who choose to vote twice and stop then according to Table~\ref{table:votedis}. Thus, she has a better success probability if, in addition to the above, when she sees a re-vote and also manipulates that, her probability of success rises to about 95.5\%. In a nutshell, for all the voter strategies of type $V\ldots V$, the probability of the pattern $V\ldots V$ is larger than the probability of $V\ldots VC$ (with the same number of $V$-s).
This means that in order to maximize the overall success probability, it is always rational for the adversary to manipulate the latest ballot in the sequence $V\ldots V$.
Note that by assuming this strategy, the attacker does not need to prepare a decision for any sequence where there are some $C$-s, since she will get caught at the corresponding verifications anyway.
Thus, the overall success probability of the optimal attacker strategy is the sum of probabilities of patterns $V$, $VV$, $VVV$, etc. Adding them for the three different events, we see that the overall probability of these is around 96\%.

From here, we can compute the attacker's success probability if she tries to implement a large-scale vote manipulation attack. E.g., the attacker wants to change $k$ votes and remain undetected. Since the probability of not getting caught when manipulating any single vote is $0.96$, the probability of remaining undetected changing $k$ votes is $0.96^k$. Consequently, the probability of being detected at least once is $1-0.96^k$.
For example, when the adversary tries to change 100 or 200 random votes, her probability of success is 1.687\% or 0.028\%, respectively.

\begin{theorem}\label{th:ucsecure}
Let $n,k,t$ be in $\poly(\lambda)$. For any $\delta = O(\log \lambda)$, the IVXV e-voting protocol $\Pivote$ described in Fig.~\ref{fig:main_protocol} UC-realizes the $\Fvote$ functionality as depicted in Fig.~\ref{fig:Fvote}   in the $\big\{ \Fbb, \mFcert, \linebreak   \{\Fvsd^{\Voter_i}\}_{i\in[n]}, \Fasd, \Fkeygen, \Fdec, \Fve^\DDD \big\}$-hybrid world against static corruption if $\mathsf{PKE}$ is IND-CPA secure, $\NIZK$ is computationally sound and composable zero-knowledge, and there is a constant fraction $0<c\leq 1$ of the voters who check their latest  submitted ballots w.r.t. the voter behaviour distribution  $\DDD$ (i.e. the voter behaviour pattern ends with `C').
\end{theorem}

\begin{proof}
%\ref{th:ucsecure}
To prove the theorem, we first construct a simulator $\SSS$ such that no non-uniform PPT environment $\ZZZ$ can distinguish with non-negligible probability between (i) the real execution $\exec^{\Fbb, \mFcert, \{\Fvsd^{\Voter_i}\}_{i\in[n]}, \Fasd, \Fkeygen, \Fdec, \Fve^\DDD}_{\Pivote,\AAA,\ZZZ}$, where the parties $\VVV:=\{\Voter_1,\ldots,\Voter_n\}$ and the trustees $\TTT:=\{\Trustee_1, \ldots, \Trustee_k\}$ run protocol $\Pivote$ in the $\big\{ \Fbb, \mFcert,\{\Fvsd^{\Voter_i}\}_{i\in[n]}, \Fasd, \Fkeygen, \Fdec, \Fve^\DDD \big\}$-hybrid world and the corrupted parties are controlled by a dummy adversary $\AAA$ who simply forwards messages from/to $\ZZZ$, and (ii) the ideal execution $\exec^{\Fbb, \mFcert}_{\Fvote,\SSS,\ZZZ}$ where the parties interact with functionality $\Fvote$ in the $\big\{ \Fbb,\mFcert \big\}$-hybrid model and corrupted parties are controlled by the simulator $\SSS$. Let  $\VVV_\cor \subseteq \VVV$ and $\TTT_\cor \subseteq \TTT$ be the set of corrupted voters and trustees, respectively. \\

\noindent \textbf{Case 1:} $ 0 \leq |\VVV_\cor| \leq n \;\wedge \; 0 \leq |\TTT_\cor| < t$
%\vspace{-5pt}
%\paragraph{\bf Simulator.}
\par
\indent{\bf Simulator.}
The simulator $\SSS$ internally runs $\AAA$, forwarding messages to/from
the environment $\ZZZ$. The simulator $\SSS$ simulates honest voters $\Voter_i \in \VVV \setminus \VVV_\cor $, honest trustees $\Trustee_j \in \TTT \setminus \TTT_\cor $ and  functionalities $\{\Fvsd^{\Voter_i}\}_{i\in[n]}, \Fasd, \Fkeygen, \Fdec$, and $\Fve^\DDD$.   In addition, the $\SSS$ simulates the following interactions with $\AAA$.
%\vspace{-5pt}
  \begin{itemize}%[\leftmargin=-2pt]
  \item In the \textbf{preparation} phase:
  \begin{itemize}
\item Upon receiving $(\Start,sid, \Trustee_j)$ from the external $\Fvote$ for an honest trustee $\Trustee_j \in \TTT \setminus \TTT_\cor$, the simulator $\SSS$ acts as $\Trustee_j$, sending message $(\Ready, sid)$ to functionality $\Fkeygen$.

\item When the simulated functionality $\Fkeygen$ receives $(\Ready, sid)$ from a corrupted trustee $\Trustee_j \in\TTT_\cor$, the simulator $\SSS$ acts as $\Trustee_j$ to send $(\Start,sid)$ to the external $\Fvote$.

\item If $\EA$ is corrupted, $\SSS$ keeps monitoring $\Fbb$; when $\PK$ is posted on the $\Fbb$, it acts as $\EA$, sending $(\Begin, sid)$ to $\Fvote$.

\item If $\EA$ is honest, upon receiving $(\Begin, sid)$ from $\Fvote$, $\SSS$ acts as $\EA$, sending $(\PubKey, sid)$ to the functionality $\Fkeygen$. When the simulated $\EA$ receives $(\PubKey, sid, \PK)$ from $\Fkeygen$, $\SSS$ acts as $\EA$, sending $(\PubKey, sid, \PK)$ to $\Fbb$.

\end{itemize}

  \item In the \textbf{voting} phase:
  \begin{itemize}

\item Upon receiving $(\Vote, sid, \Voter_i)$ from the external $\Fvote$ for an honest voter $\Voter_i \in \VVV \setminus \VVV_\cor$, the simulator $\SSS$ acts as $\Voter_i$, following the protocol  $\Pivote$ description as if $\Voter_i$ receives $(\Vote,sid,\bot)$ from $\ZZZ$.

\item The simulator $\SSS$  monitoring $\Fbb$, once a record $(ssid, c, \sigma)$ is posted on the $\Fbb$, where $ssid= (\Voter_i, ssid')$ for some corrupted voter $\Voter_i\in\VVV_\cor$, it fetches $\SK$ from the internal state of the simulated $\Fkeygen$ and compute $x\leftarrow\Dec(\SK, c)$. $\SSS$ then acts as $\Voter_i$ to send $(\Vote, sid, x)$ to $\Fvote$.

\item When the simulated $\Fvsd^{\Voter_i}$ receives $(\Corrupt, sid)$ from $\AAA$, the simulator $\SSS$ sends $(\Corrupt, sid, \Voter_i)$ to the external $\Fvote$. Once $\SSS$ receives $(\Leak, sid, \Voter_i,  x_i)$ from $\Fvote$, it acts as $\Fvsd^{\Voter_i}$ to send $(\Leak, sid, \Voter_i, x_i)$  to $\AAA$.  When the simulated $\Fvsd^{\Voter_i}$ receives $(\Tamper, sid, \TM)$ from $\AAA$: if $\TM = \emptyset$, the simulator $\SSS$
sends $(\Proceed, sid, \Voter_i, x_i)$ to $\Fvote$; otherwise, $\SSS$ computes $(y,z,\st')\leftarrow\TM(\Voter_i,\st,x_i)$. It parses $y$ as $(ssid,c,\sigma)$ and extracts $x^*\leftarrow\Dec(\SK, c)$. $\SSS$  sends $(\Proceed, sid, \Voter_i, x^*)$ to $\Fvote$.

\item If $\EA$ is honest, upon receiving $(\End, sid)$ from $\Fvote$, the simulator $\SSS$ acts as $\EA$, following the protocol  $\Pivote$ description as if $\EA$ receives $(\End,\sid)$ from $\ZZZ$.

\end{itemize}

\item In the \textbf{tally} phase:
  \begin{itemize}
  \item Upon receiving $(\Tally,sid, \Trustee_j)$ from the external $\Fvote$ for an honest trustee $\Trustee_j \in \TTT \setminus \TTT_\cor$, the simulator $\SSS$ acts as $\Trustee_j$, sending message $(\Key, sid)$ to functionality $\Fdec$.

\item When the simulated functionality $\Fdec$ receives $(\Key, sid)$ from a corrupted trustee $\Trustee_j \in\TTT_\cor$, the simulator $\SSS$ acts as $\Trustee_j$ to send $(\Tally,\sid)$ to the external $\Fvote$.

\item If $\EA$ is honest: upon receiving $(\Leak, sid, \Ballots^*)$ from $\Fvote$, the simulator $\SSS$ interprets $\Ballots^*$ as $(m_1,\ldots, m_n)$ and acts as $\Fdec$ to send the command $(\PubPost, sid, (m_1,\ldots, m_n))$ to $\Fbb$.

\item If $\EA$ is corrupted: the simulator $\SSS$ fetches $(c_1,\ldots, c_n),(c'_1,\ldots, c'_n), \pi$ from $\Fbb$ and extract $\Pi\leftarrow\NIZK.\Ext((c_1,\ldots, c_n),(c'_1,\ldots, c'_n), \pi).$
    Upon receiving the message $(\Leak, sid, \Ballots)$ from $\Fvote$, the simulator $\SSS$ interprets $\Pi(\Ballots)$ as $(m_1,\ldots, m_n)$ and act as $\Fdec$ to send the command $(\PubPost, sid, (m_1,\ldots, m_n))$ to $\Fbb$.

\item Upon receiving $(\Leak, sid, \Ballots^*)$ from $\Fvote$, the simulator $\SSS$ interprets $\Ballots^*$ as $(m_1,\ldots, m_n)$ and sends $(\PubPost, sid, (m_1,\ldots, m_n))$ to $\Fbb$.

\end{itemize}

\item In the \textbf{audit} phase:
  \begin{itemize}
  \item Upon receiving $(\Audit,sid)$ from the external $\Fvote$, the simulator $\SSS$ acts as $\AU$,  following the protocol  $\Pivote$ description as if $\AU$ receives $(\Audit, sid)$ from $\ZZZ$.
%\item Since $\AU$ is always trusted, the simulator $\SSS$ acts as $\AU$,  following the protocol  $\Pivote$ description as if $\AU$ receives $(\Audit, sid)$ from $\ZZZ$.

\end{itemize}

\end{itemize}

\noindent\textbf{Indistinguishability.} The indistinguishability is proven through a series of hybrid worlds $\HHH_0,\ldots,\HHH_2$.

%Define $\ADV_{\cA|\cZ,i,j}(1^\lambda) =\abs{ \Pr[\cA|\cZ(\cH_i)=1] - \Pr[\cA|\cZ(\cH_j)=1] }$, where $\cA|\cZ(\cH_i)$ stands for $\cZ$'s output in the experiment where $\cA$ and $\cZ$ are playing in the hybrid world $\cH_i$.

\noindent
\textbf{Hybrid $\HHH_0$:} The real protocol execution $$\exec^{\Fbb, \mFcert, \{\Fvsd^{\Voter_i}\}_{i\in[n]}, \Fasd, \Fkeygen, \Fdec, \Fve^\DDD}_{\Pivote,\AAA,\ZZZ}.$$

\noindent
\textbf{Hybrid $\HHH_1$:} $\HHH_1$ is the same as $\HHH_0$ except %the simulator uses
that $\HHH_1$: in the tally phase, $\Fdec$ does not use $\SK$ to decrypt the ciphertext $(c'_1,\ldots, c'_n)$; instead, it does as the simulator $\SSS$ as described above.

\begin{claim}
$\HHH_1$ and $\HHH_0$ are indistinguishable.

\end{claim}
%\begin{proof}
%According to Definition, the probability that $\NIZK.\Ext$ extraction fails (a.k.a. knowledge error) is negligible, so the probability that any adversary $\AAA$ and the environment $\ZZZ$ can distinguish $\HHH_1$ from $\HHH_0$ is $\negl(\secp)$.
%
%\end{proof}

\noindent
\textbf{Hybrid $\HHH_2$:} $\HHH_2$ is the same as $\HHH_1$ except %the simulator uses
that $\HHH_2$: in the voting phase, ignores the voter's choice and always use $x':=\bot$  as the voter's input.

\begin{claim}
$\HHH_2$ and $\HHH_1$ are indistinguishable if the underlying threshold public key encryption scheme is IND-CPA secure.

\end{claim}
%\begin{proof}
%We prove it by reduction. Assume there exists an adversary $\AAA$ who can distinguish $\HHH_2$ from $\HHH_1$, we can construct an adversary $\BBB$ who can break the IND-CPA game of the underlying public key encryption scheme. In particular, $\BBB$ uses the $\PK^*$ received from the IND-CPA challenger as the public key output by $\Fkeygen$. $\BBB$ then defines $M_0:=(x_1,\ldots, x_n)$ and $M_1:=(\bot,\ldots,\bot)$. The corresponding ciphertexts of $M_b$ are used as the ciphertexts generated by $\Fvsd^{\Voter_i}$. When $\AAA$ guesses $\HHH_{b^*}$, $\BBB$ outputs $b^*-1$ as its guess.  It is easy to see that if $\AAA$ can distinguish $\HHH_2$ from $\HHH_1$ with non-negligible probability then $\BBB$ can break the IND-CPA game.
%\end{proof}

%
Now let's examine the probability that $\AU$ may return different output to the environment $\ZZZ$ in the $\textbf{Hybrid $\HHH_2$}$ and the ideal execution $\exec^{\Fbb, \mFcert}_{\Fvote,\SSS,\ZZZ}$.
We will estimate the success probability of the adversary who is trying to manipulate the vote on the voter's device without getting caught.
Assume $0<c\leq 1$ fraction of the voters check their latest submitted ballots, the probability of not getting caught when manipulating any single vote is $1-c$. Therefore, to tamper $\omega(\log \secp)$ votes, the probability of being detected at least once is
$$p= 1-(1-c)^{\omega(\log \secp)} = 1 -  \negl(\secp) \enspace.$$

Therefore, with probability $p$, the adversary's view of $\HHH_2$ is identical to the simulated view of the ideal execution $\exec^{\Fbb, \mFcert}_{\Fvote,\SSS,\ZZZ}$.
 Therefore, no PPT $\ZZZ$ can distinguish the view of the ideal execution from the view of the real execution with more than negligible probability.\\

\noindent \textbf{Case 2:} $ 0 \leq |\VVV_\cor| \leq n \;\wedge \;  t \leq |\TTT_\cor| \leq k$

\smallskip
%\paragraph{\bf Simulator.}
\par\indent\textbf{Simulator.}
Similar as Case 1, the $\SSS$ internally runs $\AAA$, forwarding messages to/from the environment $\ZZZ$.
 The $\SSS$ simulates honest voters $\Voter_i\in\VVV \setminus \VVV_\cor$, honest trustees $\Trustee_j \in\TTT\setminus\TTT_\cor $ and  functionalities $\{\Fvsd^{\Voter_i}\}_{i\in[n]}, \Fasd, \Fkeygen, \Fdec$, and $\Fve^\DDD$.   In addition, the $\SSS$ simulates the following interactions with $\AAA$.

  \begin{itemize}

  \item In the \textbf{preparation} phase:

  \begin{itemize}
  \item The same as Case 1.
%\item Upon receiving $(\Start,sid, \Trustee_j)$ from the external $\Fvote$ for an honest trustee $\Trustee_j \in \TTT \setminus \TTT_\cor$, the simulator $\SSS$ acts as $\Trustee_j$, sending message $(\Ready, sid)$ to functionality $\Fkeygen$.
%
%\item When the simulated functionality $\Fkeygen$ receives $(\Ready, sid)$ from a corrupted trustee $\Trustee_j \in\TTT_\cor$, the simulator $\SSS$ acts as $\Trustee_j$ to send $(\Start,sid)$ to the external $\Fvote$.
%
%\item If $\EA$ is corrupted, $\SSS$ keeps monitoring $\Fbb$; when $\PK$ is posted on the $\Fbb$, it acts as $\EA$, sending $(\Begin, sid)$ to $\Fvote$.
%
%\item If $\EA$ is honest, upon receiving $(\Begin, sid)$ from $\Fvote$, $\SSS$ acts as $\EA$, sending $(\PubKey, sid)$ to the functionality $\Fkeygen$. When the simulated $\EA$ receives $(\PubKey, sid, \PK)$ from $\Fkeygen$, $\SSS$ acts as $\EA$, sending $(\PubKey, sid, \PK)$ to $\Fbb$.

\end{itemize}

  \item In the \textbf{voting} phase:
  \begin{itemize}

\item The same as Case 1.

\item When the simulated $\Fvsd^{\Voter_i}$ receives $(\Corrupt, sid)$ from $\AAA$, the simulator $\SSS$ sends $(\Corrupt, sid, \Voter_i)$ to the external $\Fvote$. Once $\SSS$ receives $(\Leak, sid, \Voter_i, x_i)$ from $\Fvote$, it acts as $\Fvsd^{\Voter_i}$ to send $(\Leak, sid, \Voter_i, x_i)$  to $\AAA$.  When the simulated $\Fvsd^{\Voter_i}$ receives the message $(\Tamper, sid, \TM)$ from $\AAA$: if $\TM = \emptyset$, the simulator $\SSS$
sends $(\Proceed, sid, \Voter_i, x_i)$ to $\Fvote$;
otherwise, $\SSS$ computes $(y,z,\st')\leftarrow\TM(\Voter_i,\st,x_i)$. It parses $y$ as $(ssid,c,\sigma)$ and extracts $x^*\leftarrow\Dec(\SK, c)$. $\SSS$  sends $(\Proceed, sid, \Voter_i, x^*)$ to $\Fvote$.

%otherwise, $\SSS$ monitoring the $\Fbb$. Once the corresponding $(ssid,c,\sigma)$ is posted, it parses $y$ as $(ssid,c,\sigma)$ and extracts $x^*\leftarrow\Dec(\SK, c)$. $\SSS$  sends $(\Proceed, sid, \Voter_i, x^*)$ to $\Fvote$.

\item If $\EA$ is honest, upon receiving $(\End, sid)$ from $\Fvote$, the simulator $\SSS$ acts as $\EA$, following the protocol  $\Pivote$ description as if $\EA$ receives $(\End,\sid)$ from $\ZZZ$.

\end{itemize}

\item In the \textbf{tally} phase:
  \begin{itemize}
  \item The same as Case 1.
%  \item Upon receiving $(\Tally,sid, \Trustee_j)$ from the external $\Fvote$ for an honest trustee $\Trustee_j \in \TTT \setminus \TTT_\cor$, the simulator $\SSS$ acts as $\Trustee_j$, sending message $(\Key, sid)$ to functionality $\Fdec$.
%
%
%
%\item When the simulated functionality $\Fdec$ receives $(\Key, sid)$ from a corrupted trustee $\Trustee_j \in\TTT_\cor$, the simulator $\SSS$ acts as $\Trustee_j$ to send $(\Tally,\sid)$ to the external $\Fvote$.
%
%
%\item If $\EA$ is honest: upon receiving $(\Leak, sid, \Ballots^*)$ from $\Fvote$, the simulator $\SSS$ interprets $\Ballots^*$ as $(m_1,\ldots, m_n)$ and acts as $\Fdec$ to send $(\PubPost, sid, (m_1,\ldots, m_n))$ to $\Fbb$.
%
%\item If $\EA$ is corrupted: the simulator $\SSS$ fetches $(c_1,\ldots, c_n),(c'_1,\ldots, c'_n), \pi$  from $\Fbb$ and extract $\Pi\leftarrow\NIZK.\Ext((c_1,\ldots, c_n),(c'_1,\ldots, c'_n), \pi)$.  Upon receiving $(\Leak, sid, \Ballots)$ from $\Fvote$, $\SSS$ interprets $\Pi(\Ballots)$ as $(m_1,\ldots, m_n)$ and act as $\Fdec$ to send $(\PubPost, sid, (m_1,\ldots, m_n))$ to $\Fbb$.
%
%\item Upon receiving $(\Leak, sid, \Ballots^*)$ from $\Fvote$, the simulator $\SSS$ interprets $\Ballots^*$ as $(m_1,\ldots, m_n)$ and sends $(\PubPost, sid, (m_1,\ldots, m_n))$ to $\Fbb$.

\end{itemize}

\item In the \textbf{audit} phase:
  \begin{itemize}
  \item The same as Case 1.
%  \item Upon receiving $(\Audit,sid)$ from the external $\Fvote$, the simulator $\SSS$ acts as $\AU$,  following the protocol  $\Pivote$ description as if $\AU$ receives $(\Audit, sid)$ from $\ZZZ$.

\end{itemize}

\end{itemize}

\noindent\textbf{Indistinguishability.}
The indistinguishability in this case is straightforward, as $\SSS$ never simulate a single message to either any corrupted parties or the external $\Fbb$. The simulator $\SSS$ knows all the honest voters' ballot from $\Fvote$ when $|\TTT_\cor|\geq t$.
Meanwhile, it also extracts the ballot of the malicious voters by using $\SK$ leaked from $\Fkeygen$. Hence, the simulator $\SSS$ can submit the extracted ballot to the external $\Fvote$ on the malicious voters' behave. Moreover, the $\EA$ cannot tamper any ballots without being detected due to the unforgeability of the underlying signature scheme modeled by $\mFcert$. Therefore, the view of $\ZZZ$ in the ideal execution has identical distribution to the view of $\ZZZ$ in the real execution.
%\qed
\end{proof}

\section{Real-world Instantiation }\label{sec:implementation}
%{Realization of Ideal Functionalities}\label{sec:realization}
%! TEX encoding =UTF-8 Unicode
%! TEX root = mainArxiv.tex

%\subsection{The Estonian IVXV Voting System Implementation}\label{sec:implementation}

\begin{figure}[t!]

    \centering
    \includegraphics[width=0.48\textwidth]{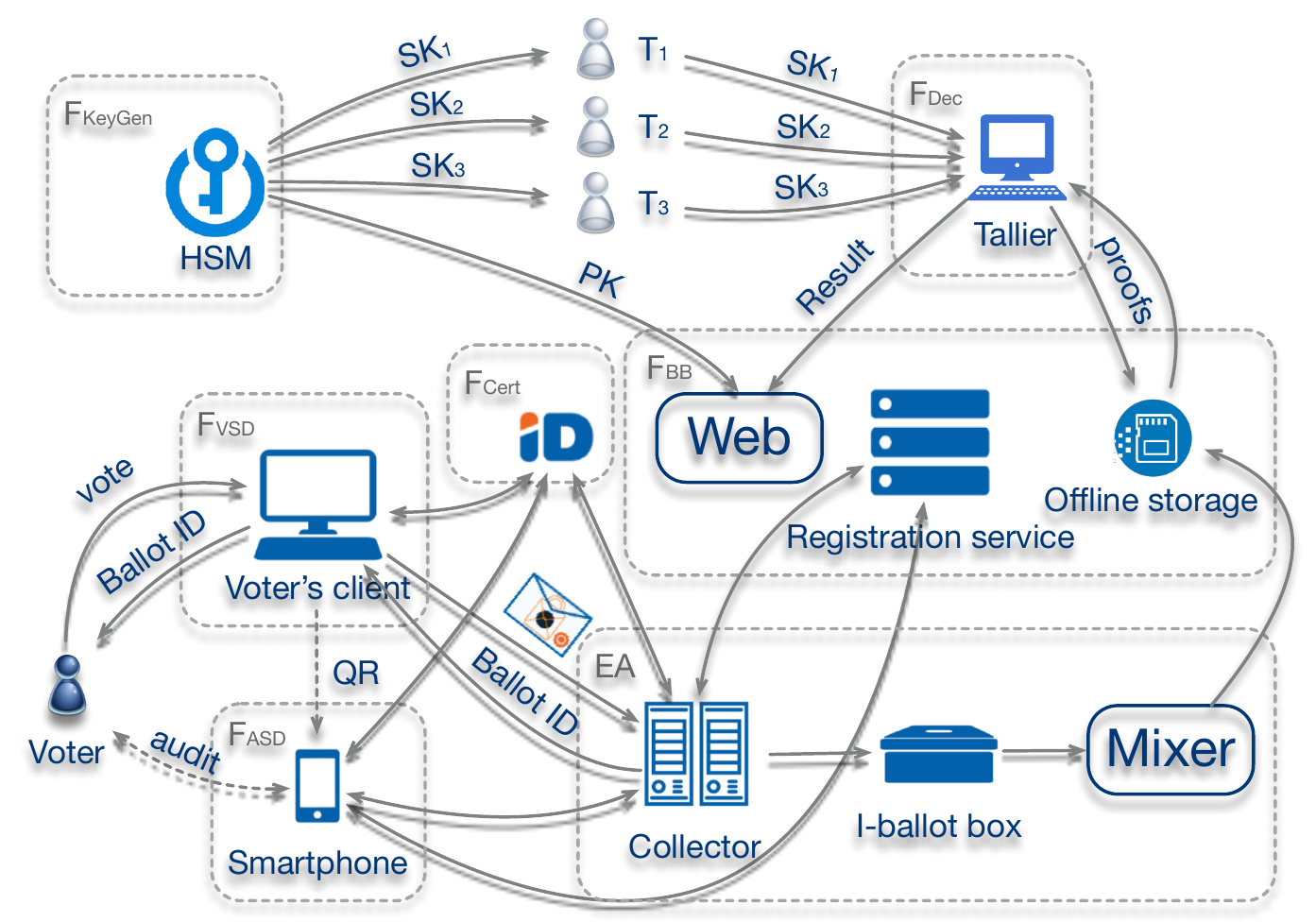}
    \caption{IVXV real-world protocol overview}
    \label{fig:ivxv}
\end{figure}

In this section, we examine how those ideal functionalities and system entities are instantiated in the real-world Estonian IVXV voting system as described in the official documentation~\cite{IVXV}. Before that, we first provide a brief description. The Estonian i-voting process is depicted in Fig.~\ref{fig:ivxv}, and it consists of three stages.

\vspace{5pt}
\noindent\textbf{Pre-voting stage.}
Before the election, the lists of candidates and eligible voters are prepared. Also, the client software for i-voting is compiled and distributed via the main i-voting website (\url{https://www.valimised.ee/en}). This software will serve as a front-end for eligibility verification, downloading the candidate list, and formatting the voter choice as a vote.

IVXV uses Elgamal encryption as the underlying re-randomizable public-key encryption scheme, and the election key is generated and distributed through an offline ceremony. $k$ (human) trustees are selected from the members of \emph{national electoral committee} and   \emph{state electoral office}. During the ceremony, the trustees are gathered on-site and witness the \emph{hardware security module} (HSM) generating a public and secret key pair $(\PK,\SK)$, and splitting the secret key $\SK$ into $k$ shares  using $(t,k)$ Shamir secret sharing as $(\SK_1,\ldots, \SK_9)\leftarrow\Deal(\SK)$. The shares are then stored on $k$ smartcards, one for each trustee. $\SK$ is then deleted from the HSM. The values $k=9,t=5$ are used in recent Estonian i-voting practice.

\vspace{5pt}
\noindent\textbf{Voting stage.}
The voters are authenticated by their physical eID tokens (ID chip card or Mobile-ID SIM card) combined with a knowledge-based PIN. During the voting stage, the voter first downloads the list of candidates specifies his/her choice in the i-voting software on his/her \emph{client}; the client then encrypts the vote with a fresh random coin $r$ and signs the ciphertext. It then sends the signed ciphertext to the \emph{collector}. After checking the validity of the signature, the collector sends the ciphertext to the \emph{registration service} who will record it and sign it together with a time stamp. The collector then generates a unique ballot ID and sends it back to the voter's client. After that, it stores the doubly-signed ciphertext (indexed by the ballot ID) in the \emph{i-ballot box}.  In the end, the client outputs a QR code that contains the random coin $r$ and the ballot ID.

\begin{figure}[htbp!]

    \centering
    \includegraphics[width=0.45\textwidth]{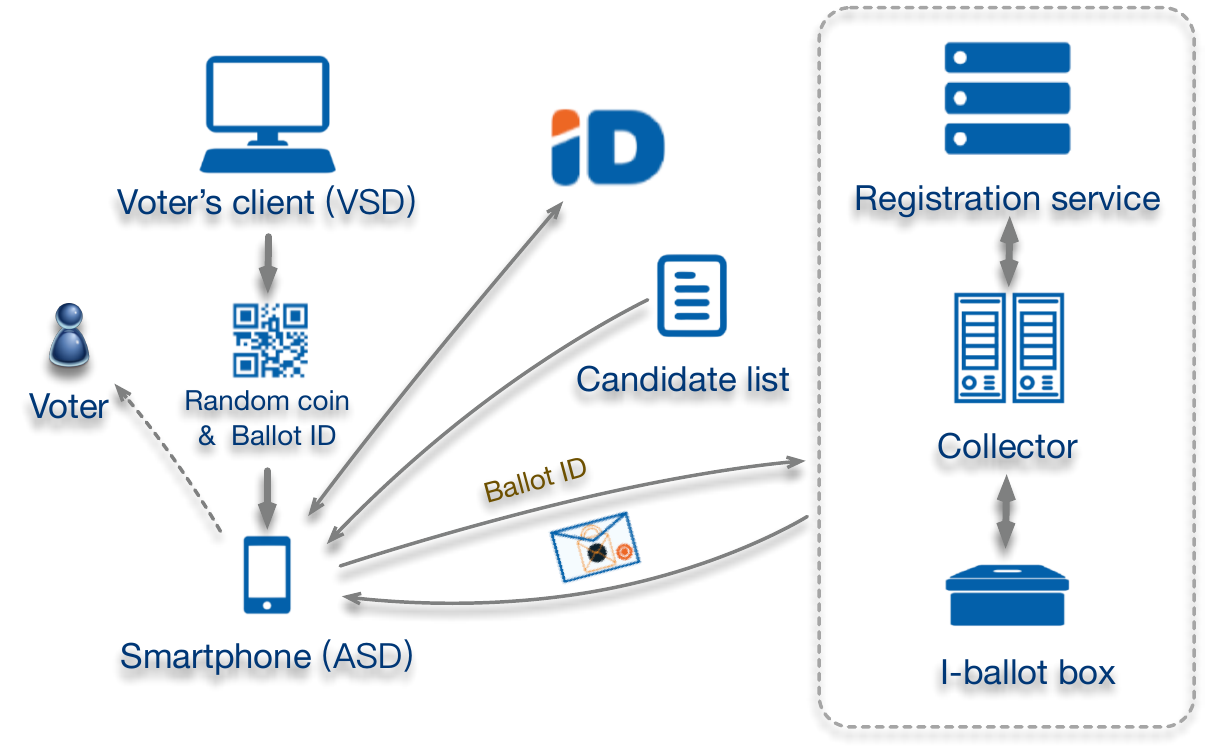}
    \caption{Voter ballot verification}
    \label{fig:verify}
\end{figure}

If the voter would like to audit the ballot, he/she can use a smartphone (with the verification app) and scan the QR code obtaining the random coin $r$ and ballot ID\footnote{The ballot can only be verified within one hour after submission.}. As shown in Fig.~\ref{fig:verify}, during the verification,  the smartphone queries the collector and i-ballot box with the ballot ID, obtaining the doubly-signed ciphertext. The smartphone first checks the validity of the registration service's signature and the voter's signature. Let the ElGamal ciphertext be $c:=(c_1,c_2)$ and $\PK:=(g,h)$. It then computes $m:=\TDec(\PK, c,r) = c_2\cdot h^{-r}$ and  outputs $m$ as the candidate number to the voter for confirmation on the mobile device screen. The voter will then have to make the decision if $m$ matches his/her true intent.

\vspace{5pt}
\noindent\textbf{Processing stage.}
At the end of the election, the ballots stored in the i-ballot box are sent to the \emph{mixer}, who will drop the attached signatures and shuffle the ballots with a NIZK proof. Those shuffled ballots are then decrypted and counted in an offline ceremony. During the ceremony, the trustees are gathered on-site and submit their key shares via smartcards to a \emph{tallier} machine. The tallier combines the private key and decrypts the ciphertexts one by one with NIZK proofs.
 The tally result is counted, and invalid votes are discarded. The final election result is announced on the main i-vote website, and all the NIZK proofs are kept on an offline medium, available to all the third-party auditors.

\vspace{5pt}
\noindent\textbf{Entity/Functionality mapping.}

\vspace{3pt}
\noindent\underline{$\EA$.} The election authority $\EA$ consists of the collector, the i-ballot box, and the mixer.

\vspace{3pt}
\noindent\underline{$\mFcert$.} The (multi-session) certificate functionality $\mFcert$ can be instantiated with the readily deployed Estonian national PKI infrastructure, i.e. ID-card, Mobile-ID and Digi-ID. Since the digital identification service is not exclusively setup for i-voting, we model $\mFcert$ as a global functionality in the gUC \cite{TCC:CDPW07} framework.

\vspace{3pt}
\noindent\underline{$\Fvsd$.} The $\Fvsd$ functionality is instantiated by the voter's client, the i-voting software, as well as the national ID card reader for signing operations.

\vspace{3pt}
\noindent\underline{$\Fasd$.} The $\Fasd$ functionality can be instantiated by any smartphone with the verification app installed.

\vspace{3pt}
\noindent\underline{$\Fbb$.} The main i-vote website can be viewed as the public BB, while the registration service and the offline storage medium (which records the shuffled ballot and their decryption NIZK proofs) can be viewed as the private BB.

\vspace{3pt}
\noindent\underline{$\Fkeygen$.}  The $\Fkeygen$ is instantiated as a combination of cryptographic, physical, and organizational measures in the form of a key generation ceremony.

\vspace{3pt}
\noindent\underline{$\Fdec$.}  Similarly, the auditable decryption functionality $\Fdec$ is instantiated as a combination of cryptographic, physical, and organizational measures in the form of a voting ceremony. The subtle difference from the key generation ceremony is that it uses an untrusted tallier for decryption. Nevertheless, if the taller cheated during the decryption process, the auditors can detect it by checking the corresponding decryption NIZK proofs.

\section{Related works}\label{sec:relatedUC}% on UC-modeling of e-voting systems}

 In terms of modeling e-voting as a ceremony, in 2017, Kiayias \emph{et al.}~\cite{DBLP:conf/pkc/Kiayias0Z17} analyzed the security of Helios in terms of an e-voting ceremony using property-based definitions.
%
% there are only a few works can be found in the literature
%
%
%most of existing UC-secure voting schemes such as \cite{ACNS:Groth04,VOTEID:MarPerQui07,DTISSEC:MorNao10,E:SzePre15} assume that they have an anonymous broadcast channel, some public keys are set up, and voters are registered without specifying how this is done. In more detail,
In 2004, Groth \cite{ACNS:Groth04} suggested evaluating voting schemes in the UC framework, and the ideal functionality corresponds closely to the well-known ballot box model used today in manual voting. In this case, security properties such as privacy, accuracy, and robustness follow as easy corollaries.
In order to achieve an End-to-end auditable voting system along with the security properties such as correctness, privacy, fairness, and receipt-freeness, etc., Marneffe, Pereira, and Quisquater \cite{VOTEID:MarPerQui07} investigated the use of techniques from the simulation-based security tradition for the analysis of these existing protocols, through a case-study on the ThreeBallot protocol. Further, Moran and Naor~\cite{DTISSEC:MorNao10} presented a split-ballot voting scheme with the ``everlasting privacy'' property, and formally proven the security of the proposed protocol in the universal composability framework, based on number-theoretic assumptions. Alwen et al.~\cite{C:AOZZ15} proposed the first
construction of a UC receipt-free e-voting protocol, which implies that
 the first construction of an MPC protocol (for more than two parties) that
is incredibly secure and universally composable.
%E:SzePre15
Afterward, Szepieniec and Preneel \cite{E:SzePre15} presented a novel unifying framework for electronic voting in the UC model that includes universal verifiability.
\section{Conclusion}\label{sec:conclusion}

%! TEX encoding =UTF-8 Unicode
%! TEX root = mainArxiv.tex

To the best of our knowledge, the Estonian IVXV with end-to-end verifiability may now be the only country allowing access to Internet voting to all its citizens. Although the Estonia IVXV can provide the available performance for the user, but there is no formal systematical security analysis of the current Estonian IVXV.
In this paper, we analyzed the security of the Estonian
IVXV with the help ceremony under the UC framework.
Furthermore,
we also present the details of the protocol, the results of implementation, and discuss the performance thereof according to Estonian Internet voting statistics collected from the recent 5-year legally binding national-wide elections.

%\section*{Acknowledgment}
%The authors are grateful to the  anonymous reviewers for their invaluable comments that improve  the completeness and readability of this paper.

{
\def\shortbib{0}
{\scriptsize
%\begin{spacing}{0.9}

 \bibliographystyle{IEEEtran}
%\scriptsize
\bibliography{FinancialCrypto19}
%\end{spacing}
}}

%\input{CV}
%\appendix
%\appendices
%\input{appendix}
\end{document}